%% file: maintext.tex
\documentclass[runningheads]{llncs}
\usepackage{graphicx}
\usepackage[margin=1in]{geometry}
\usepackage{caption}
\usepackage{subcaption}
\usepackage{hyperref}
\usepackage{dirtytalk}
\usepackage{wrapfig}
\usepackage{algorithm, algpseudocode}

\usepackage{xcolor}

\usepackage[symbol]{footmisc}

\newcommand{\hide}[1]{}

\DeclareMathAlphabet{\mathpzc}{OT1}{pzc}{m}{it}

\newtheorem{notation}{Notation}

\usepackage{amssymb,color,epsf}
\usepackage{mathtools}
\usepackage{enumerate}

\usepackage{bm}
\usepackage{amsmath}
\usepackage{tabularx}
\usepackage{color, colortbl}
\usepackage{url}

\DeclareMathAlphabet{\mathpzc}{OT1}{pzc}{m}{it}

\usepackage [cmtip,arrow]{xy}
\xyoption{all}
\usepackage {pb-diagram,pb-xy}
\usepackage[mathscr]{eucal}
\usepackage[utf8]{inputenc}
\usepackage[T1]{fontenc}
\usepackage[english]{babel}
\usepackage{bm}
\usepackage{tabularx}
\usepackage{bm}

\usepackage{blkarray, bigstrut}
\usepackage{gauss}

\newcommand {\junk}[1]{}

\newcommand {\R} {\mathbb{R}}

\newcommand {\RR} {{\mathcal R}}

\newcommand {\la}   {{\langle}}
\newcommand {\ra}   {{\rangle}}

\newcommand {\PP}     {\mathbb{P}} %projective space

\def\addots{\mathinner{\mkern1mu
\raise1pt\vbox{\kern7pt\hbox{.}}
\mkern2mu\raise4pt\hbox{.}\mkern2mu
\raise7pt\hbox{.}\mkern1mu}}

\newcommand{\HH}  {\mbox{\rm H}}

\DeclareMathOperator{\diag}{diag}

\newcommand{\x}{\mathbf{x}}

\newcommand{\y}{\mathbf{y}}

\newcommand{\vb}{\mathbf{v}}

\newcommand{\tb}{\mathbf{t}}

\newcommand{\Ob}{\mathrm{Ob}}

\newcommand{\nc}{\newcommand}
\newcommand{\rc}{\renewcommand}
\nc{\mc}{\mathcal}
\rc{\t}{\text}
\nc{\op}[1]{\operatorname{#1}}
\nc{\opcat}[1]{\mathbf{#1}}

\nc{\id}{\op{id}}

\nc{\umutnote}[1]{{\marginpar{\small \textcolor{blue}{#1}}}}

\nc{\cA}{\mc{A}}\nc{\cB}{\mc{B}}\nc{\cC}{\mc{C}}\nc{\cD}{\mc{D}}\nc{\cE}{\mc{E}}\nc{\cF}{\mc{F}}\nc{\cG}{\mc{G}}\nc{\cH}{\mc{H}}\nc{\cI}{\mc{I}}\nc{\cJ}{\mc{J}}\nc{\cK}{\mc{K}}\nc{\cL}{\mc{L}}\nc{\cM}{\mc{M}}\nc{\cN}{\mc{N}}\nc{\cO}{\mc{O}}\nc{\cP}{\mc{P}}\nc{\cQ}{\mc{Q}}\nc{\cR}{\mc{R}}\nc{\cS}{\mc{S}}\nc{\cT}{\mc{T}}\nc{\cU}{\mc{U}}\nc{\cV}{\mc{V}}\nc{\cW}{\mc{W}}\nc{\cX}{\mc{X}}\nc{\cY}{\mc{Y}}\nc{\cZ}{\mc{Z}}

\rc{\PP}{\mathbb{P}}
\rc{\AA}{\mathbb{A}}
\nc{\bbC}{\mathbb{C}}
\nc{\CC}{\mathbb{C}}

\nc{\code}[1]{{\texttt{#1}}}
\nc{\mcode}[1]{{\text{\texttt{#1}}}}
\nc{\xto}[1]{\raisebox{-0.03cm}{\scalebox{0.85}{$\,\xrightarrow{#1}\,$}}}
\nc{\xtonormal}[1]{\xrightarrow{#1}}
\nc{\xfrom}[1]{\xleftarrow{#1}}
\nc{\sidenote}[1]{\marginpar{\small #1}}

\nc{\Aff}{\opcat{Aff}}
\nc{\AffVar}{\opcat{AffVar}}
\nc{\ProjVar}{\opcat{ProjVar}}
\nc{\GAP}{\opcat{GrAlgPairs}}
\nc{\GA}{\opcat{GrAlg}}

\nc{\acc}{\mathrm{a.c.c}}
\nc{\GL}{\mathrm{GL}}

\nc{\Mod}{\t{-}\opcat{Mod}}
\nc{\Sub}{\opcat{Sub}}
\nc{\iso}{\cong}
\nc{\compose}{\circ}

\newcommand{\bp}{\begin{proof}}
\newcommand{\ep}{\end{proof}}

\newcommand{\econt}{\mathrm{content}}
\begin{document}
\title{
Probing omics data via harmonic persistent homology
%Harmonic rapresentation provide insighs of omics data.
\thanks{Supported by IBM Research}
}

\author{Davide Gurnari\inst{1}\thanks{These authors contributed equally to this work}\orcidID{0000-0002-3668-8711}
 \and
Aldo Guzmán-Sáenz\inst{2}$^\dagger$\orcidID{0000-0003-2725-621X}
 \and
Filippo Utro\inst{2}\orcidID{0000-0003-3226-7642}
\and
Aritra Bose\inst{2}\orcidID{0000-0002-8665-056X}
\and
Saugata Basu\inst{3}\orcidID{0000-0002-2441-0915}
\and
Laxmi Parida\inst{2}\orcidID{0000-0002-7872-5074}
}
\authorrunning{D. Gurnari et al.}
\institute{Dioscuri Centre in Topological Data Analysis, Mathematical Institute PAN, Warsaw, PL \and
IBM  Research, Yorktown Heights, NY, USA\\
\and
Dept. of Mathematics, Purdue University, West Lafayette, IN, USA
\\
\email{parida@us.ibm.com}}
\maketitle              % typeset the header of the contribution

\begin{abstract}
Identifying molecular signatures from complex disease patients with underlying symptomatic similarities is a significant challenge in the analysis of high dimensional multi-omics data. Topological data analysis (TDA) provides a way of extracting such information from the geometric structure of the data and identifying multiway higher-order relationships. Here, we propose an application of \textit{Harmonic} persistent homology, which overcomes the limitations of ambiguous assignment of the topological information to the original elements in a representative topological cycle from the data. When applied to multi-omics data, this leads to the discovery of hidden patterns highlighting the relationships between different omic profiles, while allowing for common tasks in multi-omics analyses, such as disease subtyping, and most importantly biomarker identification for similar latent biological pathways that are associated with complex diseases. Our experiments on multiple cancer data show that \textit{harmonic} persistent homology effectively dissects multi-omics data to identify biomarkers by detecting representative cycles predictive of disease subtypes.  

\end{abstract}
%
%
%
%\clearpage
\section{Introduction}
Omics studies have gained substantial importance in unraveling interactions between biomarkers that underlie complex diseases given the increasing availability of such data from across modalities due to recent technological advances~\cite{hasin2017multi}. These biomarkers are instrumental in clinical decision-making and drug discovery. However, extracting them from complex data can often be a challenge with the high dimensionality of these datasets as well as the heterogeneous molecular profiles of patients, disease subtypes, and other biases that plague epidemiological studies~\cite{libbrecht2015machine}. Furthermore, identifying biomarkers that share similar underlying biological pathways or molecular profiles among patients can be an even bigger impediment that still need investigation and novel analytical tools. One of the most robust approaches that analyzes data with a new perspective is Topological Data Analysis (TDA). TDA has been shown as a useful tool for the analysis of omics data~\cite{basu_et_al:LIPIcs:2018:9316,bose_cuna_2021,liao2019tmap,zheng2021hidef,vipond2021multiparameter,platt_characterizing_2016,platt_epidemiological_2022,10.1371/journal.pone.0126383,mcguirl2020topological,xia2014persistent}, and it enables the identification of (complex) multiway high order relationships in the data. 

Persistent homology is a tool from TDA used to study the topology of spaces at different scales with algebraic constructs. This approach has been used  successfully  in the context of pattern discovery (see for example ~\cite{basu_et_al:LIPIcs:2018:9316}). One limitation of this approach is an inherent ambiguity when mapping back the information 
obtained from the topology of a space (say the geometric realization of a simplicial complex)  
to the basic elements (in this case the vertices of the simplicial complex). 
In this paper we overcome this problem by applying a new tool introduced in \cite{basu2022harmonic} -- namely  \emph{Harmonic} Persistent Homology. 
Use of harmonic persistent homology yields an unambiguous way of selecting representative cycles 
from homology classes, which furthermore satisfy certain theoretical guarantees (simplices, which are essential for a class, appear with highest weight). Use of harmonic homology thus overcomes
a basic obstacle for the use of TDA in the context of genomic data analysis, where identifying 
important simplices (which encodes biologically relevant relationships) is crucial. Although TDA and persistent homology have been applied in analyzing multi-omics data~\cite{liao2019tmap,vipond2021multiparameter,zheng2021hidef}, they were different in scope as they simply used persistent homology to identify broad spatial patterns in multi-omics data and were plagued by the aforementioned issue of mapping the learned topological structure back to the elements or vertices in the simplicial complex. We perform computation of persistent harmonic cycles with our software library, called \texttt{\textbf{maTilDA}}, available at \href{https://github.com/IBM/matilda}{https://github.com/IBM/matilda}.

% One of the main analysis done using TDA is via the study of bar diagram and pattern discovery on them~\cite{basu_et_al:LIPIcs:2018:9316}. One of the main caveat of the TDA is to going back to the original data.  Persistent homology is a tool from TDA used to study the topology of spaces at different scales with algebraic constructs. One limitation of this approach is that there is an inherent ambiguity when mapping back the information of the topology of a space to its elements. A possible way to overcome this limitation is through, the notion of harmonic persistent homology was introduced in~\cite{basu_harmonic}, which in particular yields an unambiguous way to select representative cycles associated to bars satisfying certain mild conditions: If we assume that bars are \emph{simple}, then we can select a representative cycle in a well defined manner. Intuitively, a bar is simple if it is the only bar with those exact start and end times. Note that, in our case, we can assume that all bars are simple since we can find a sufficiently small perturbation of the original point cloud with its VR complex having all bars simple. We computation of persistent harmonic cycles with our software library, called maTilDA, available at \href{https://github.com/IBM/matilda}{https://github.com/IBM/matilda}.

\begin{figure}[!htbp]
    \centering
    \includegraphics[width=0.95\linewidth]{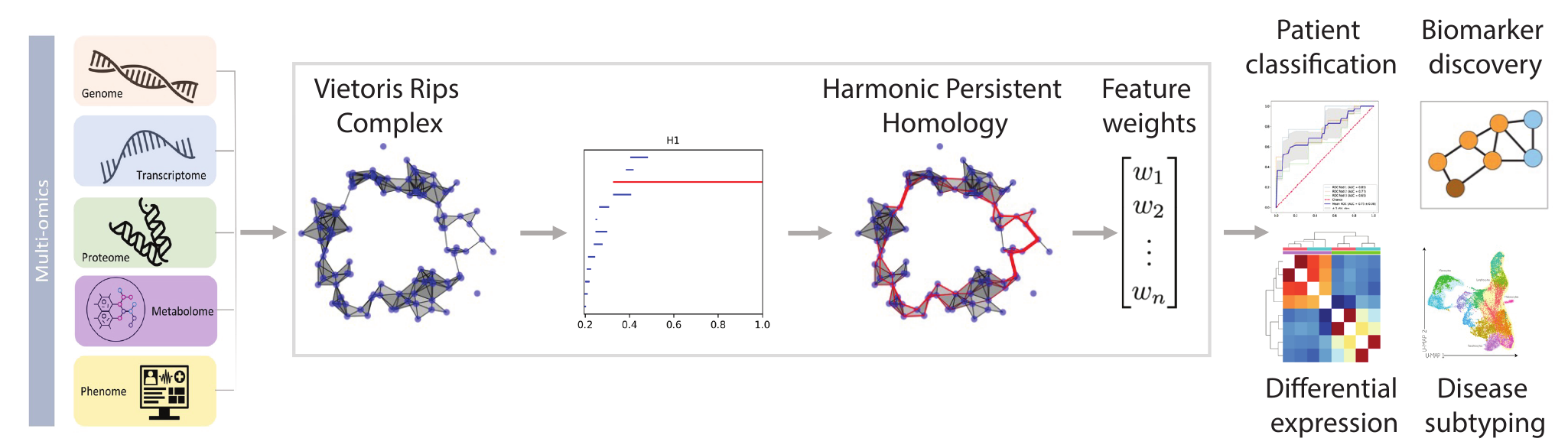}
    \caption{Schematic of the Harmonic Framework. Given a set of multi-omic samples, the algorithm builds a topological structure on top of the input data according to a predefined similarity measure; and computes its persistent homology barcode. A subset of bars is selected from the barcode and their corresponding harmonic representatives are computed. The harmonic representative corresponding to the red bar in the diagram is depicted in red in \textit{Harmonic Persistent Homology }panel, edges' thickness are proportional to their harmonic weights. Harmonic weights on the simplices (e.g. edges) can then be converted to weights on the samples and used as feature vectors for a series of downstream tasks (we limit ourselves to a few examples).}
    \label{fig:workflow}
\end{figure}

Intuitively, harmonic persistent homology establishes relationships between features or observations in the data that may lead to the discovery of hidden patterns and/or novel insights owing to the capability of TDA to analyze data at different scales. Moreover, harmonic persistent homology is naturally equipped to analyze high order interactions between data points, enabling explorations beyond simple pairwise interactions by analyzing homological features in dimensions higher than one. A schematic of our framework is given in Figure~\ref{fig:workflow}. In this manuscript, we show how harmonic persistent homology brings to light novel relationships and deeper comprehension of biological data that can help generate new hypotheses using multi-omics data. Harmonic persistent homology has the ability to assign weights on the simplices and therefore in the feature space that it spans. These weights can be used to scale the distribution of multi-omics variables providing an additional information on importance of those variables. Due to the absence of methods and data that addresses the precise problem, it is hard to perform a  benchmark comparison of the different tools reported in literature. Therefore, in this manuscript, we demonstrate the capability of  harmonic persistent homology by reproducing and expanding on known results from the literature. In particular, using different datasets, such as bulk RNA from CLL patients~\cite{10.1182/blood.2022016600}, single cell RNA (scRNA) data from Richter Syndrome (RS) and CLL patients ~\cite{Parry2023-bp}, and multi-omics lung adenocarcinoma (LUAD) and breast cancer (BRCA) data from The Cancer Genome Atlas (TCGA) database\footnote{https://www.cancer.gov/tcga}, we validate the patterns found by our framework using external clinical knowledge and expected baseline patterns.
\paragraph{Roadmap.} In the next section we give the mathematical underpinnings for the essential
simplices and harmonic persistent homology. In Section~\ref{sec:Results}, we
apply our framework to different datasets and summarize the results.

\section{Methods}
The goal of this section is to provide the mathematical background for our Harmonic Framework. In Section~\ref{subsec:simplicial} we recall some basic definitions from simplicial homology theory. The interested reader is directed to~\cite{edelsbrunner_computational_2022} for a more in-depth background on Topological Data Analysis. We define Harmonic Homology in Section~\ref{subsec:harmonic_chains} and introduce Harmonic Persistent Homology in Section~\ref{subsec:phh}. We explain the relation between essential simplices and the harmonic representative in Section~\ref{subsec:essential} and provide a toy example in Figure~\ref{fig:barcode_example} (and Section 1 in Supplementary Note). Finally, we describe how to apply this framework to real-world data in Section~\ref{subsec:harmonic_from_data}, with a specific use-case in Algorithm~\ref{alg:hph_subypes}.

% \subsection{Harmonic Persistent Homology}
% \label{subsec:harmonic_persistent_homology}

\subsection{Simplicial complex and boundary maps}
\label{subsec:simplicial}
\begin{definition}
\label{def:simplicial-complex}
A \emph{finite simplicial complex} $K$ is a set of ordered subsets of $[N] = \{0,\ldots, N\}$ for some $N \geq 0$, such that if $\sigma \in K$ and $\tau$
is a subset of $\sigma$, then $\tau \in K$. 
\end{definition}
\vspace*{-\baselineskip}
\begin{notation}
\label{def:p-simplices}
If $\sigma = \{i_0,\ldots,i_p\} \in K$, 
with $K$ a finite simplicial complex,
and $i_0 < \cdots < i_p $, we will denote $\sigma = [i_0,\ldots,i_p]$ and call $\sigma$ a \emph{$p$-dimensional simplex of $K$}. We will denote by $K^{(p)}$ the set of $p$-dimensional simplices of $K$.
\end{notation}

\begin{definition}[Chain groups and their standard bases]
\label{def:chain-groups}
Suppose $K$ is a finite simplicial complex.
For $p \geq 0$,  we will denote by 
$C_p(K) = C_p(K,\R)$ (the $p$-th chain group),   
the $\R$-vector space generated by the elements of $K^{(p)}$,
i.e.
\[
C_p(K) = \bigoplus_{\sigma \in K^{(p)}}\R \cdot \sigma.
\]
The tuple
$\left(\sigma \right)_{\sigma \in K^{(p)}}$ is then a basis (called the standard basis) of $C_p(K)$
(where by a standard abuse of notation we identify $\sigma$ with the image of $1\cdot \sigma$ under the canonical injection of the direct summand $\R \cdot \sigma$ into $C_p(K) = \bigoplus_{\sigma \in K^{(p)}}\R \cdot \sigma$).
\end{definition}
\begin{definition}[The boundary map]
\label{def:boundary-map}
We denote by $\partial_p(K): C_p(K) \rightarrow C_{p-1}(K)$ the 
linear map (called the $p$-th \emph{boundary map})  defined as follows. Since 
$\left(\sigma \right)_{\sigma \in K^{(p)}}$ is a basis of $C_p(K)$ it is enough to define the image of each $\sigma \in C_p(K)$. 
We define for $\sigma = [i_0,\ldots,i_p] \in K^{(p)}$,
\[
\partial_p(K)(\sigma) = \sum_{0\leq j \leq p} (-)^j [i_0,\ldots, \widehat{i_j}, \ldots,i_p] \in C_{p-1}(K),
\]
where $\widehat{\cdot}$ denotes omission.
\end{definition}

One can easily check that the boundary maps $\partial_p$ satisfy the key property that 
\[
\partial_{p+1}(K) \;\circ\; \partial_{p}(K) = 0,
\]
or equivalently
that 
\[
\mathrm{Im}(\partial_{p+1}(K)) \subset \ker(\partial_p(K)).
\]

\begin{notation}[Cycles and boundaries]
We denote 
\[
Z_p(K) = \ker(\partial_p(K)),
\]
(the space of \emph{$p$-dimensional cycles})
and
\[
B_p(K) = \mathrm{Im}(\partial_{p+1}(K))
\]
(the space of \emph{$p$-dimensional boundaries}).
\end{notation}

\begin{definition}[Simplicial homology groups]
\label{def:simplicial-homology}
The \emph{$p$-dimensional simplicial homology group $\HH_p(K)$} is defined as \[
\HH_p(K) = Z_p(K)/B_p(K).
\]
\end{definition}
Note that $\HH_p(K)$ is a finite dimensional $\R$-vector space.

\subsection{Representing homology classes by harmonic chains} 
\label{subsec:harmonic_chains}

Let $K$ be a finite simplicial complex.
We make the chain group $C_p(K)$ into an Euclidean space by fixing an inner product defined by:

\begin{equation}
\label{eqn:standard}
\la \sigma, \sigma'\ra = \delta_{\sigma,\sigma'},  \sigma,   \sigma' \in K^{(p)}
\end{equation}
(i.e. we declare the basis  $\left(\sigma \right)_{\sigma \in K^{(p)}}$ to be an orthonormal basis).

\begin{definition}[Harmonic homology subspace]
\label{def:harmonic}
For $p \geq 0$,  we will denote 
\[
\mathfrak{h}_p(K) = Z_p(K)  \cap  B_p(K)^\perp.
\]
and call $\mathfrak{h}_p(K) \subset C_p(K)$ the \emph{harmonic homology subspace of $K$}. 
%%(Note that $\mathfrak{h}_p(K)$ depends on the chosen inner product on $C_p(K)$.)
\end{definition}

In terms of matrices,  we have the following description of $\mathfrak{h}_p(K)$ as a subspace of $C_p(K)$.
For $p \geq 0$, let $\mathcal{A}_p(K)$ denote an orthonormal basis  of $C_p(K)$ 
(for example, if the chosen inner product is the standard one given in \eqref{eqn:standard}, then we can take
$\mathcal{A}_p(K) = \{\sigma | \sigma \in K^{(p)}\}$). Let $M_p(K)$  denote the matrix
of $\partial_p$ with respect to the basis $\mathcal{A}_p(K)$ of $C_p(K)$,  
and the basis $\mathcal{A}_{p-1}(K)$ of $C_{p-1}(K)$.
Then,  
$\mathfrak{h}_p(K)$ can be identified as the subspace of  $C_p(K)$  which is  is equal to
the intersection of the nullspaces of the two matrices $M_p(K)$ and $M_{p+1}(K)^T$.  More precisely,
\begin{equation}
\label{eqn:matrices}
  z \in \mathfrak{h}_p(K)  \Leftrightarrow [z]_{\mathcal{A}_p(K)}  \in \mathrm{null}(M_p(K)) \cap \mathrm{null}(M_{p+1}(K)^T).
\end{equation}

%%sb provisionally
%%%%%%%%%%%%%%%%%%%%%%%%%%%%%%%%%%%%%%%%%%%%%%%%%%%%%%%%%%%%%%%%%%%%%%%%%%%%%%%%%%%%%%%%

The dimensions of the harmonic homology subspaces defined above coincide with the dimensions
of the corresponding simplicial homology groups. In fact there is a well known 
canonically defined isomorphism between the two. We include the proof this well known fact below for completeness.

\begin{proposition}
\label{prop:f}
The map  $\mathfrak{f}_p(K)$ defined by 
\begin{equation}
    \label{eqn:f}
    z + B_p(K) \rightarrow \mathrm{proj}_{B_p(K)^\perp}(z), z \in Z_p(K)
\end{equation}
gives an isomorphism
$
\mathfrak{f}_p(K): \HH_p(K)  \rightarrow \mathcal{H}_p(K).
$
\end{proposition}

\begin{proof}
First observe that 
using 
the fact that $B_p(K) \subset Z_p(K)$,
we have that for $z \in Z_p(K)$, $\mathrm{proj}_{B_p(K)^\perp}(z) \in Z_p(K)$, and so the
map $\mathfrak{f}_p(K)$ is well defined. The injectivity and surjectivity of $\mathfrak{f}_p(K)$ are
then obvious.
\end{proof}
The harmonic homology group
$\mathcal{H}_p(K)$ 
is also equal to the kernel of the linear map 
$\Delta_p = \partial_{p+1}\circ \partial_{p+1}^* + \partial_p^*\circ \partial_p$.
We omit the proof of this classical fact (see for example, \cite{basu2022harmonic-journal}).
The linear map $\Delta_p(K):C_p(K) \rightarrow C_p(K)$ is a discrete analog of the Laplace operator and 
thus it makes sense to call its kernel the space of harmonic cycles.

\subsection{Persistent Harmonic Homology}
\label{subsec:phh}

Even though the data that we deal with in this paper does not have a geometric origin, 
in order to visualize the concept of persistent homology  it is useful to first consider
the case of a point-cloud $S\subseteq\mathbb{R}^n$ that approximates some underlying manifold $M$. Since we only have access to $S$, with it we will  construct an object that approximates the \say{shape} of $M$. To do this, consider $n$-dimensional balls of radius $r$ centered on elements of $S$ and denote their union by $X_r$. A well-known result in algebraic topology known as the \emph{Nerve Lemma} assures us that the \say{shape} of the union of these balls is encoded in a simplicial complex (Definition~\ref{def:simplicial-complex}) constructed using intersections of the aforementioned balls (more specifically, we consider the \emph{nerve} of the covering formed by the balls). The choice of radius $r$ is not clear, however, and thus we let it take all values in $[0,\infty)$ while keeping track of \say{shape} changes in $X_r$.

As one can easily visualize, as $r$ starts growing from $0$, there are many spurious homology classes that are born and quickly die off (e.g. the corresponding holes are filled in) and these have nothing to do with the topology of $M$. Persistent homology is a tool that can be used to separate this ``noise'' from the  \emph{bona fide} homology classes of $M$. 
\hide{
The persistent homology of the filtration $\mathcal{F}$ is encoded as a set of intervals (called bars) in the barcode of the filtration $\mathcal{F}$ (see Definition~\ref{def:barcode2})).
Intervals (bars) of short length corresponds to noise, while the ones which
are long (persistent) reflect the homology of the underlying manifold $M$.
The barcode of the filtration associated to $X$ can be used as a feature of $X$ for learning or comparison purposes. In particular, the barcodes of two 
finite sets $X, X'$, which are ``close'' as finite metric spaces, are themselves close under an appropriately defined notion of distance between barcodes. 
}

The mathematical notion that encapsulates these intuitions is called \emph{Persistent Homology}. 
For a finite filtered simplicial complex $X$, that is, a sequence of finite simplicial complexes $\{X_i\}_{1\leq i \leq n}$ with  $X_i\subseteq{X_j}$ if $i\leq j$ taking only a finite number of different values, we define their \emph{Persistent Homology} of degree $p$ by $$PH_p(X) = (\{\HH_p(X_i)\}_{1\leq i \leq n},\{\iota^{i,j}_{*}\}_{i,j \in \{1,\ldots,n\}})$$ and their \emph{Persistent Homology Groups}  by
\[
\HH^{i,j}_p(X) = \iota^{i,j}_{*}(X_p)
\]
where $\iota^{i,j}_{*}$ is the morphism in homology induced by the inclusion $X_i\subseteq{X_j}$. 

The structure of $PH_p(X)$ is that of a \emph{persistence module} (which is a graded module
over the graded ring of polynomials in one variable). 
%%A now standard result in persistent homology allows us to summarize the information contained in %%$PH_d(X)$ through the order of its generators under certain conditions. 
%%More specifically, if $X$ is a finite filtered simplicial complex and $H_*$ is homology with field coefficients $\mathbb{F}$, then 
There exists a finite multiset $D=\{(r_1,s_1),\ldots,(r_l,s_l)\}$ of pairs satisfying $r_i<s_i$ 
that encodes the isomorphism class of $PH_p(X)$. The pairs $(r_i,s_i)$ have interpretations
as the birth-time and death-time of homology classes that appear in the various 
$\HH_p(X_i)$, and is usually called (and displayed as) 
the \emph{barcode} of the filtration $\mathcal{F}$ (the pairs $(r_i,s_i)$ themselves are referred to as bars of the filtration).

Bars of short length corresponds to noise, while the ones which
are long (persistent) reflect the homology of the underlying manifold $M$.
The barcode of the filtration associated to $S$ can be used as a feature of $S$ for learning or comparison purposes. In particular, the barcodes of two 
finite sets $S, S'$, which are ``close'' as finite metric spaces, are themselves close under appropriately defined notion of distance between barcodes. An example of a filtration with its corresponding barcode is shown in Figure \ref{fig:example_filtration} (and Section 1 in Supplementary Note).

Notice that the barcode of a filtration  (being just a set of ordered pairs of numbers) do not record 
any information about the underlying simplicial complex -- in particular, about the structure and the
cycles representing the homology classes whose births and deaths are encoded by the bars.
Adding that information is not a trivial task. Strictly speaking homology classes representing bars are themselves elements of certain sub-quotients of vector spaces spanned by the simplices of the 
complex -- and this introduces an inherent ambiguity in choosing representative cycles to represent bars. The authors of \cite{basu2022harmonic} overcome this problem by introducing an extra structure 
(inner product) on the vector spaces (the so called chain spaces) arising in the definition of homology -- and using this inner product there is a canonical choice of cycles (called harmonic cycles) which represent the bars in the persistent bar codes.

It turns out that the harmonic
bars have a certain additional property (proved in \cite{basu2022harmonic}) which makes them suitable choice as representatives. The harmonic bars put higher weights on simplices that are \emph{essential} -- meaning that the simplices that \emph{must} appear with non-zero coefficient in \emph{any} choice of representative cycle for that particular bar. This last property indicates that the weight of the simplices in harmonic representatives of the bars in a bar diagram of filtrations coming from genomic data could carry important information about biological relationships. The main goal of this paper is to understand how far this is true by applying harmonic persistence theory to multi-omics data sets and check whether it can perform certain tasks, like developing prognostic biomarker signatures in complex diseases prognosis, predicting subtypes and discovering biomarkers for accelerating therapeutic discovery.

\hide{
and such that  we have an isomorphism of persistence modules
\[
PH_d(X) \cong \bigoplus_{(r,s)\in D} M^{r,s},
\]
where $M^{r,s}$ is a persistence module defined by $M^{r,s}_i = \mathbb{F}$ if $r\leq i\leq s$ and $0$ otherwise. Such collection $D$ of pairs is known as the \emph{Persistence Diagram} or, when interpreted as a collection of intervals, the \emph{Barcode} of $X$.
}

\begin{figure}
\centering
\begin{subfigure}{.4\textwidth}
\begin{subfigure}{.25\textwidth}
    \centering
    \includegraphics[width=\textwidth]{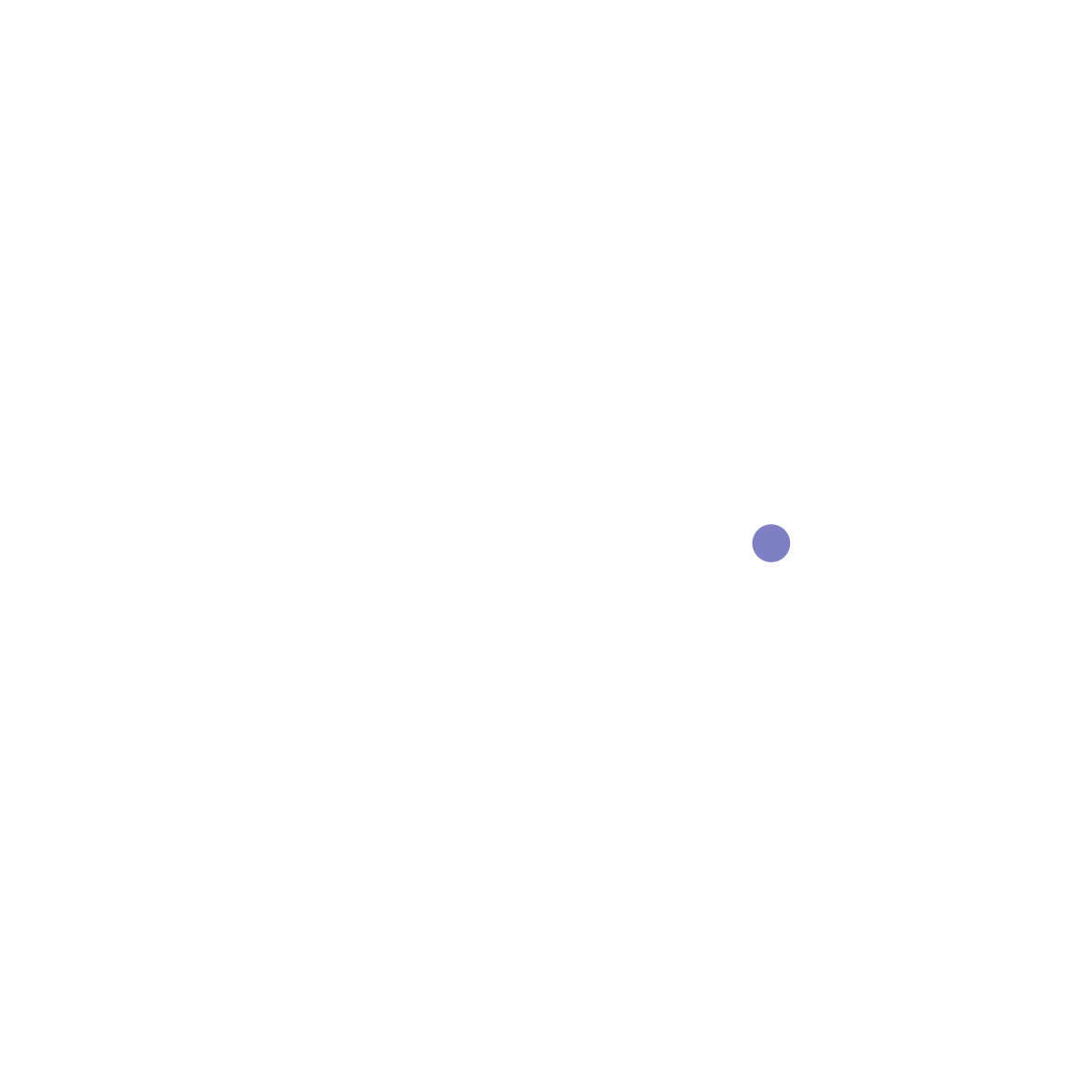}
    \subcaption{t=0}\label{fig:t0}
\end{subfigure}
\begin{subfigure}{.25\textwidth}
    \centering
    \includegraphics[width=\textwidth]{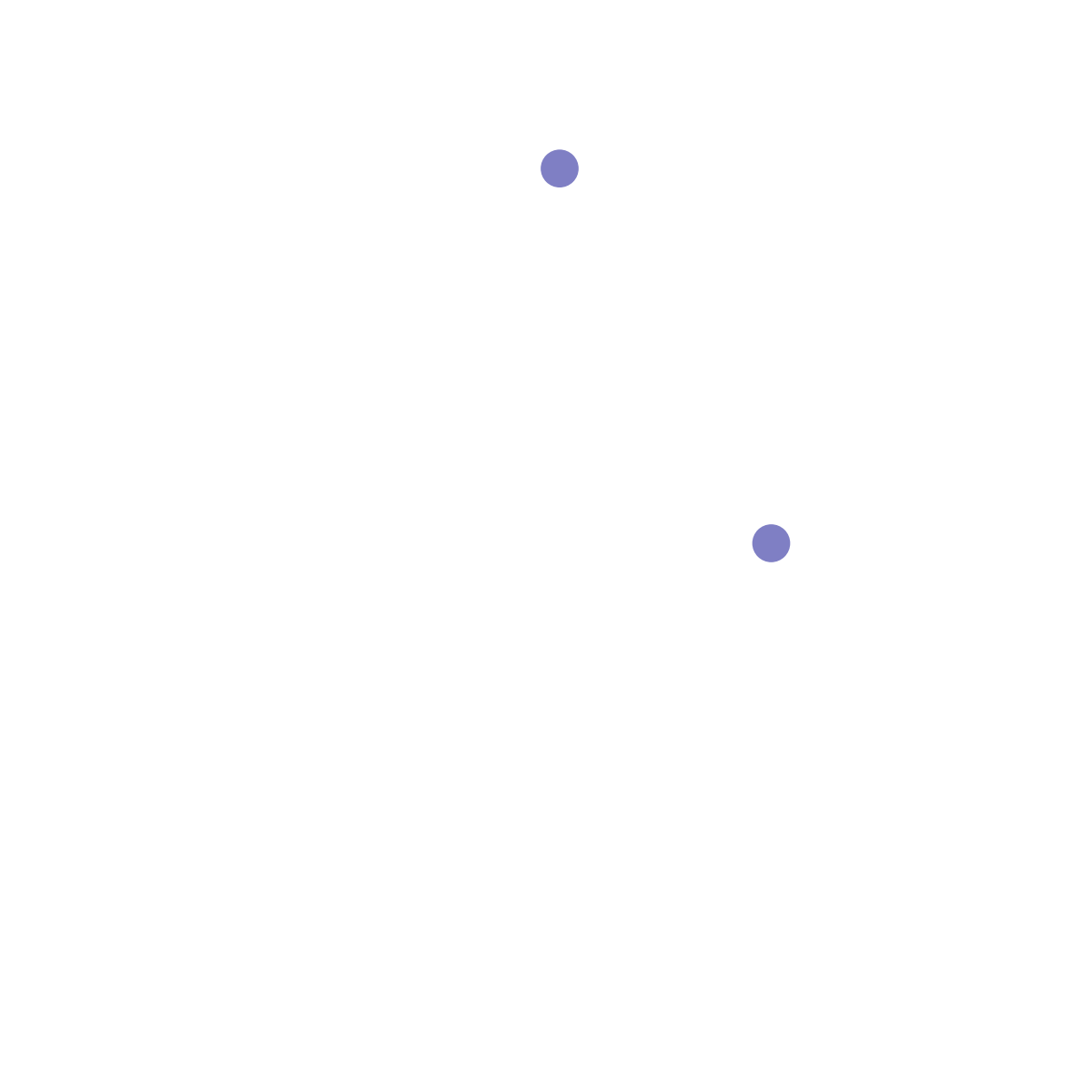}
    \subcaption{t=1}
\end{subfigure}
\begin{subfigure}{.25\textwidth}
    \centering
    \includegraphics[width=\textwidth]{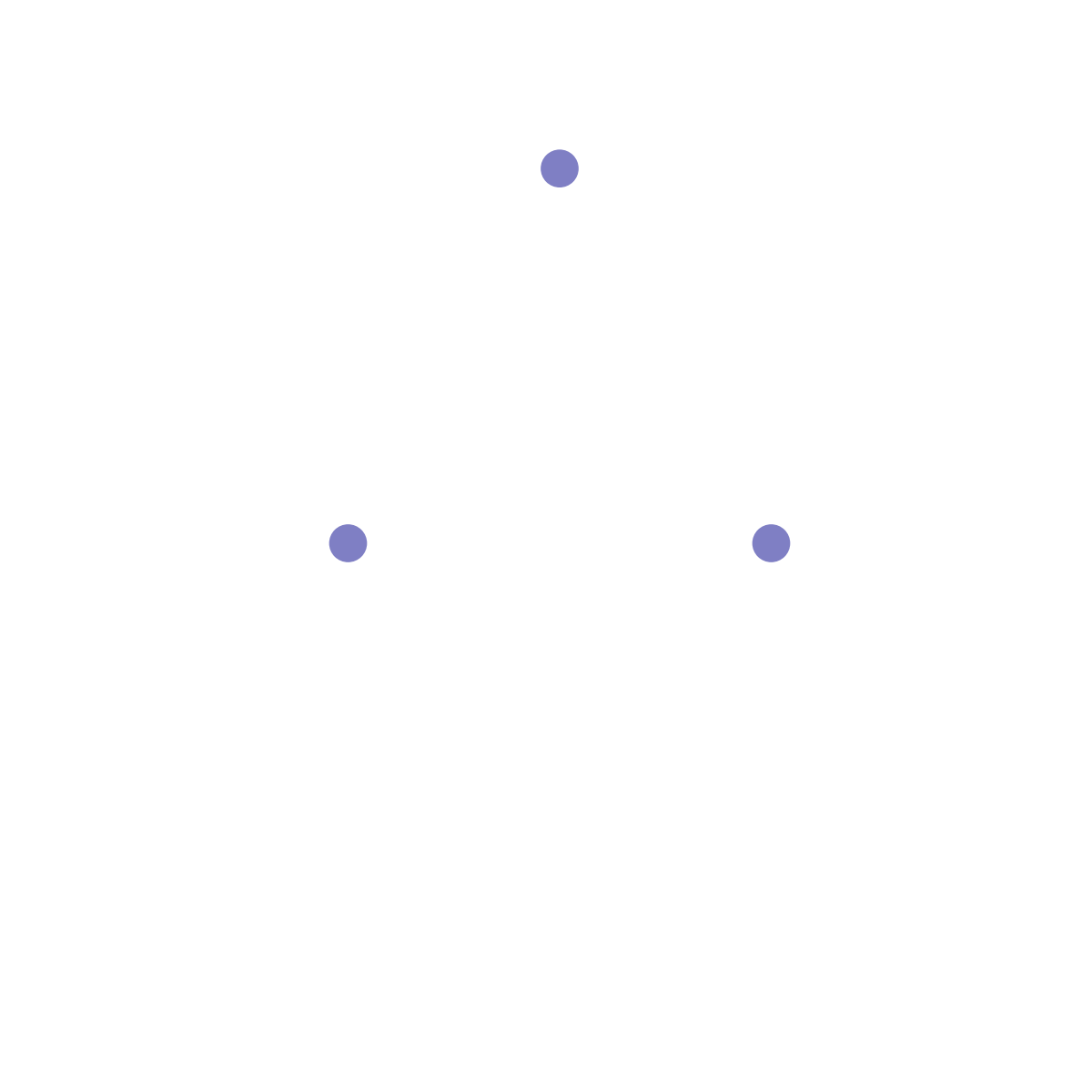}
    \subcaption{t=2}
\end{subfigure}

\begin{subfigure}{.25\textwidth}
    \centering
    \includegraphics[width=\textwidth]{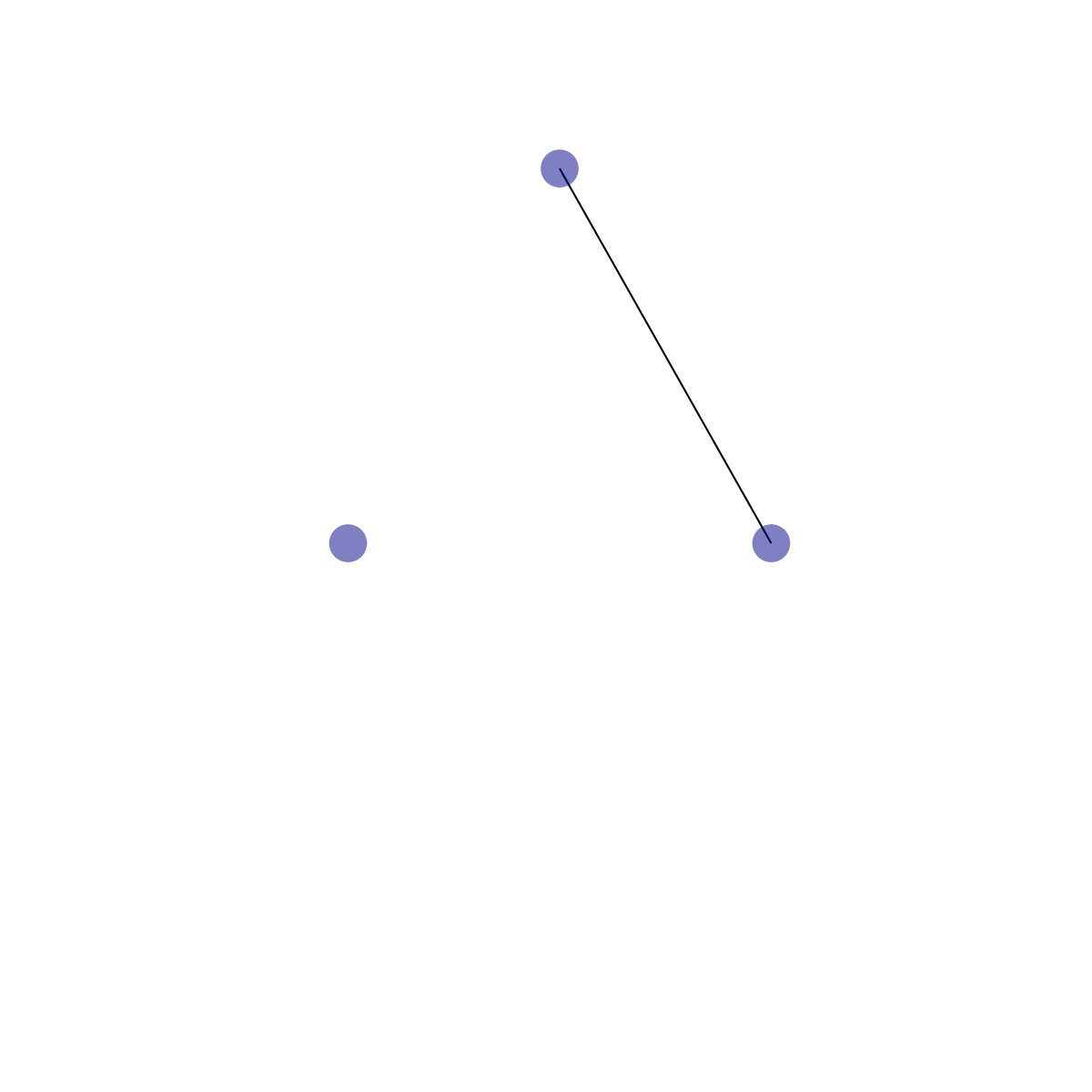}
    \subcaption{t=3}
\end{subfigure}
\begin{subfigure}{.25\textwidth}
    \centering
    \includegraphics[width=\textwidth]{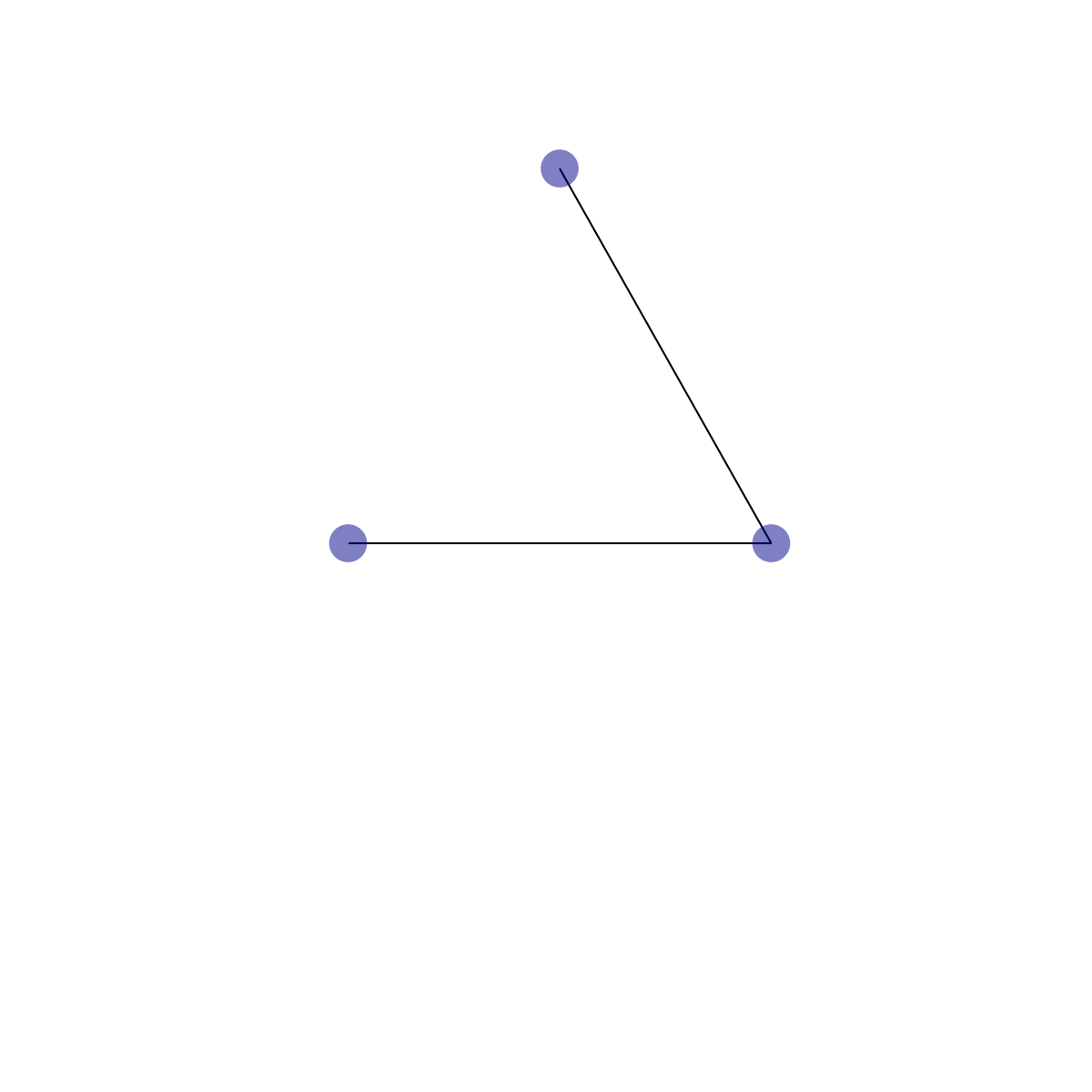}
    \subcaption{t=3}
\end{subfigure}
\begin{subfigure}{.25\textwidth}
    \centering
    \includegraphics[width=\textwidth]{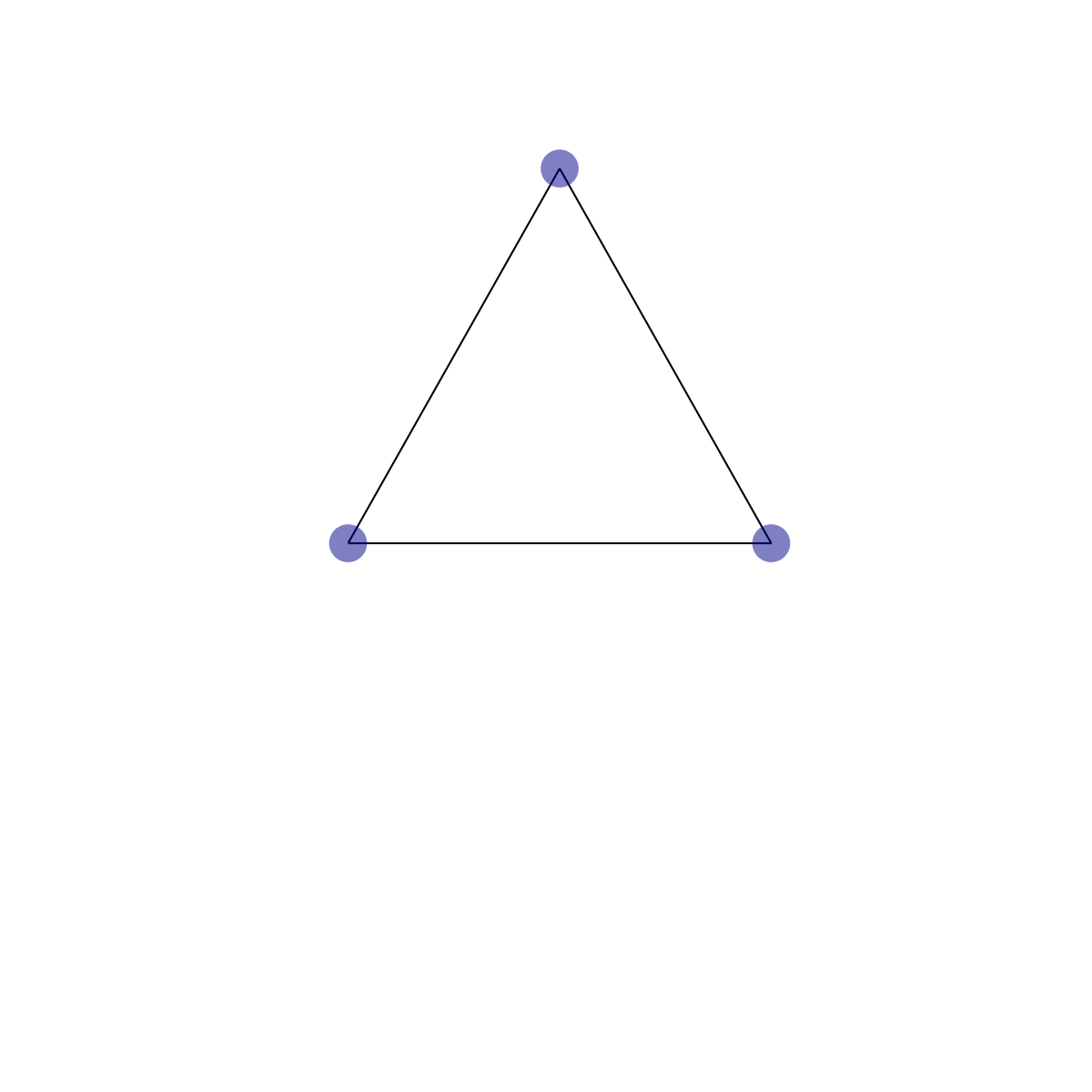}
    \subcaption{t=3}
\end{subfigure}

\begin{subfigure}{.25\textwidth}
    \centering
    \includegraphics[width=\textwidth]{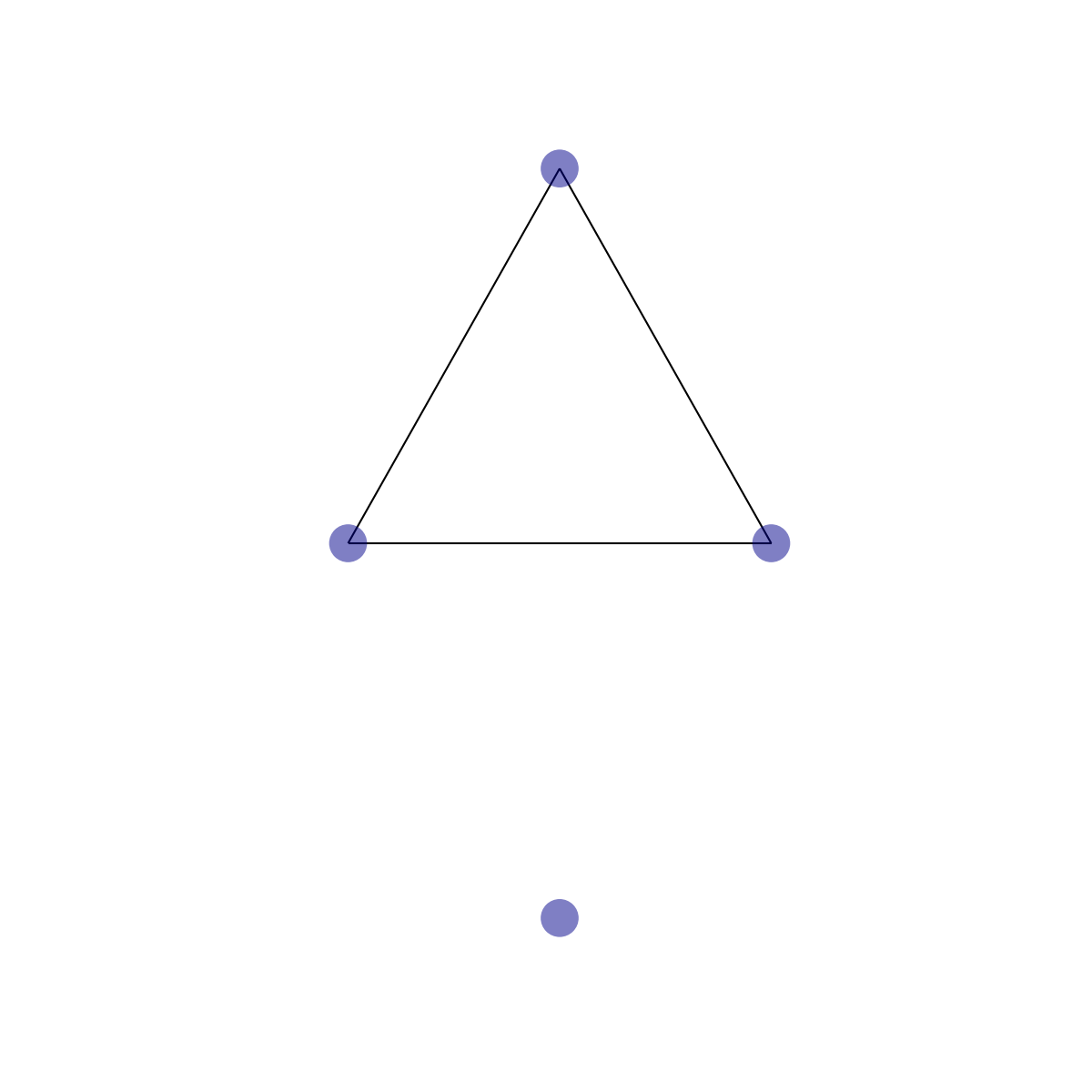}
    \subcaption{t=4}
\end{subfigure}
\begin{subfigure}{.25\textwidth}
    \centering
    \includegraphics[width=\textwidth]{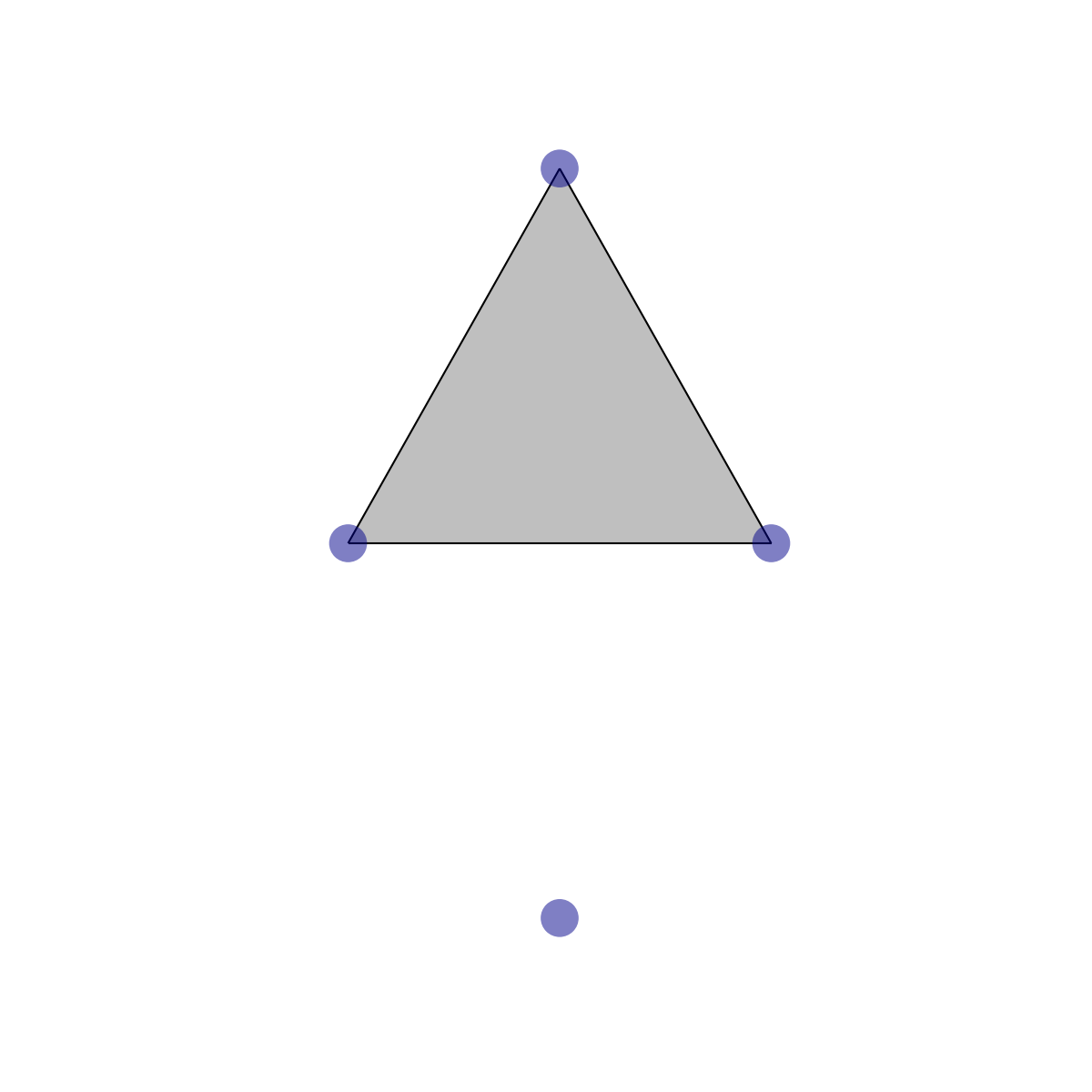}
    \subcaption{t=5}
\end{subfigure}
\begin{subfigure}{.25\textwidth}
    \centering
    \includegraphics[width=\textwidth]{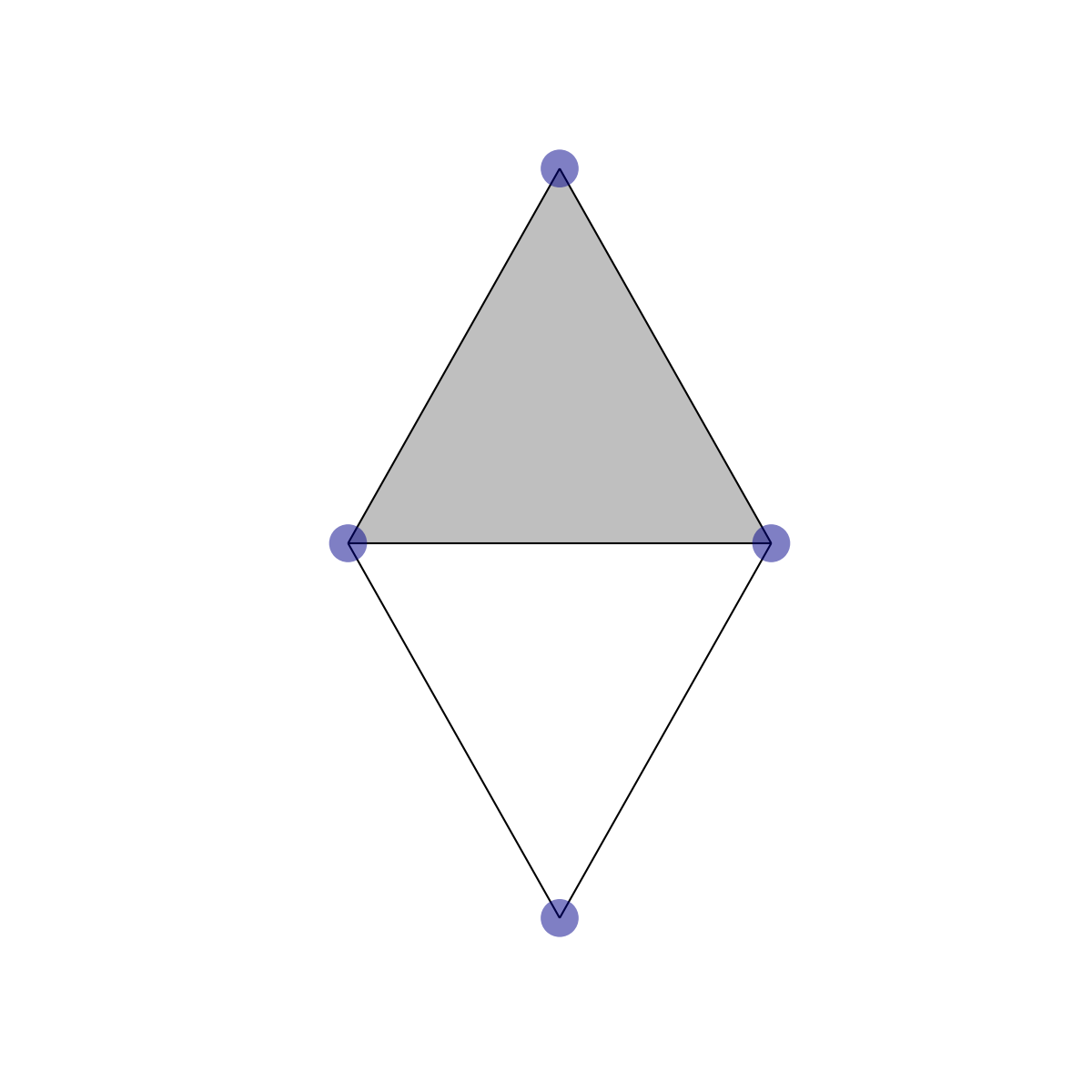}
    \subcaption{t=6}\label{fig:t6}
\end{subfigure}

\end{subfigure}
\begin{subfigure}{.3\textwidth}
\centering
\begin{subfigure}{\textwidth}
    \centering
    \includegraphics[width=\textwidth]{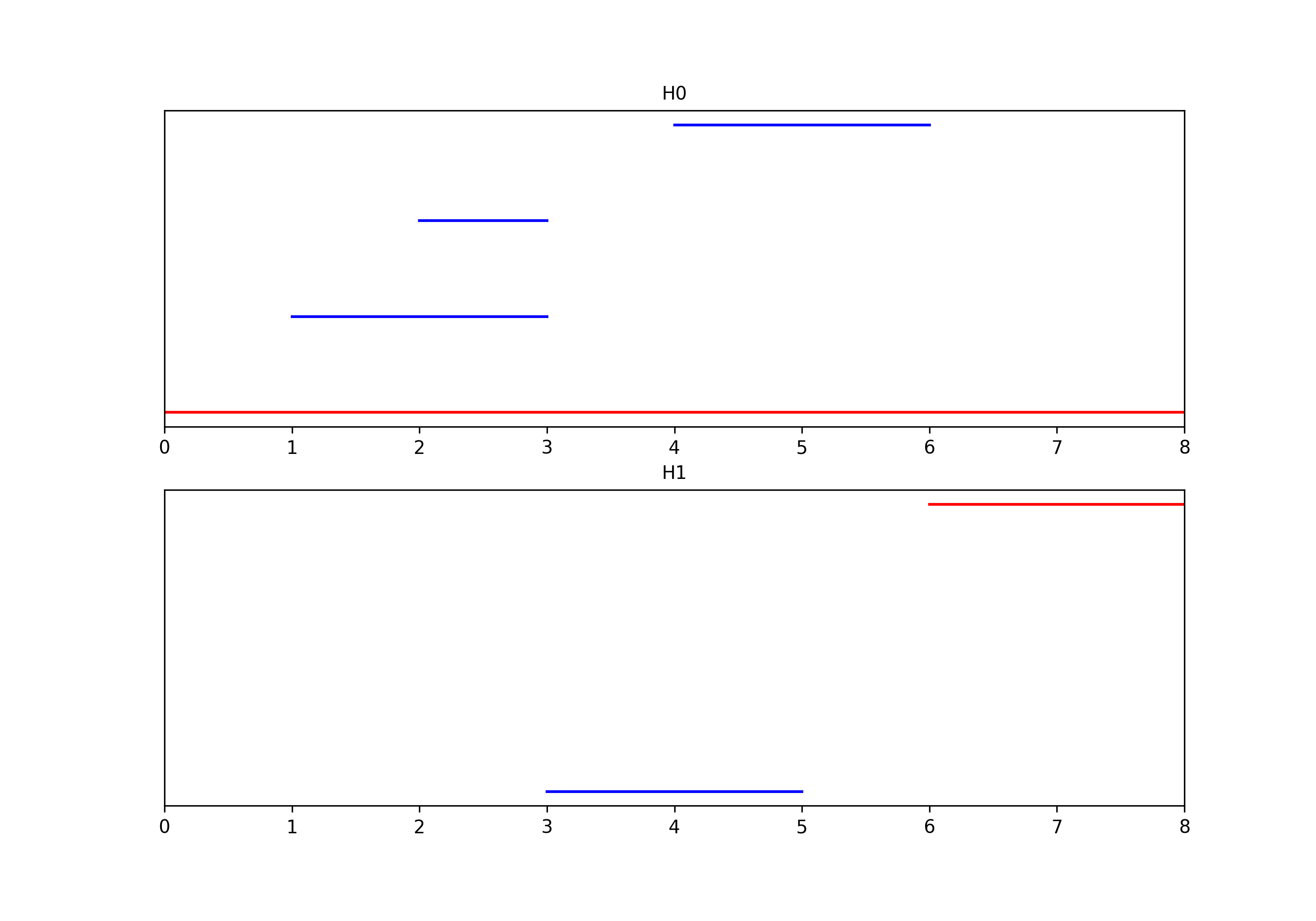}
    \subcaption{}\label{fig:barcode_example}
\end{subfigure}
\begin{subfigure}{.3\textwidth}
    \centering
    \includegraphics[width=\textwidth]{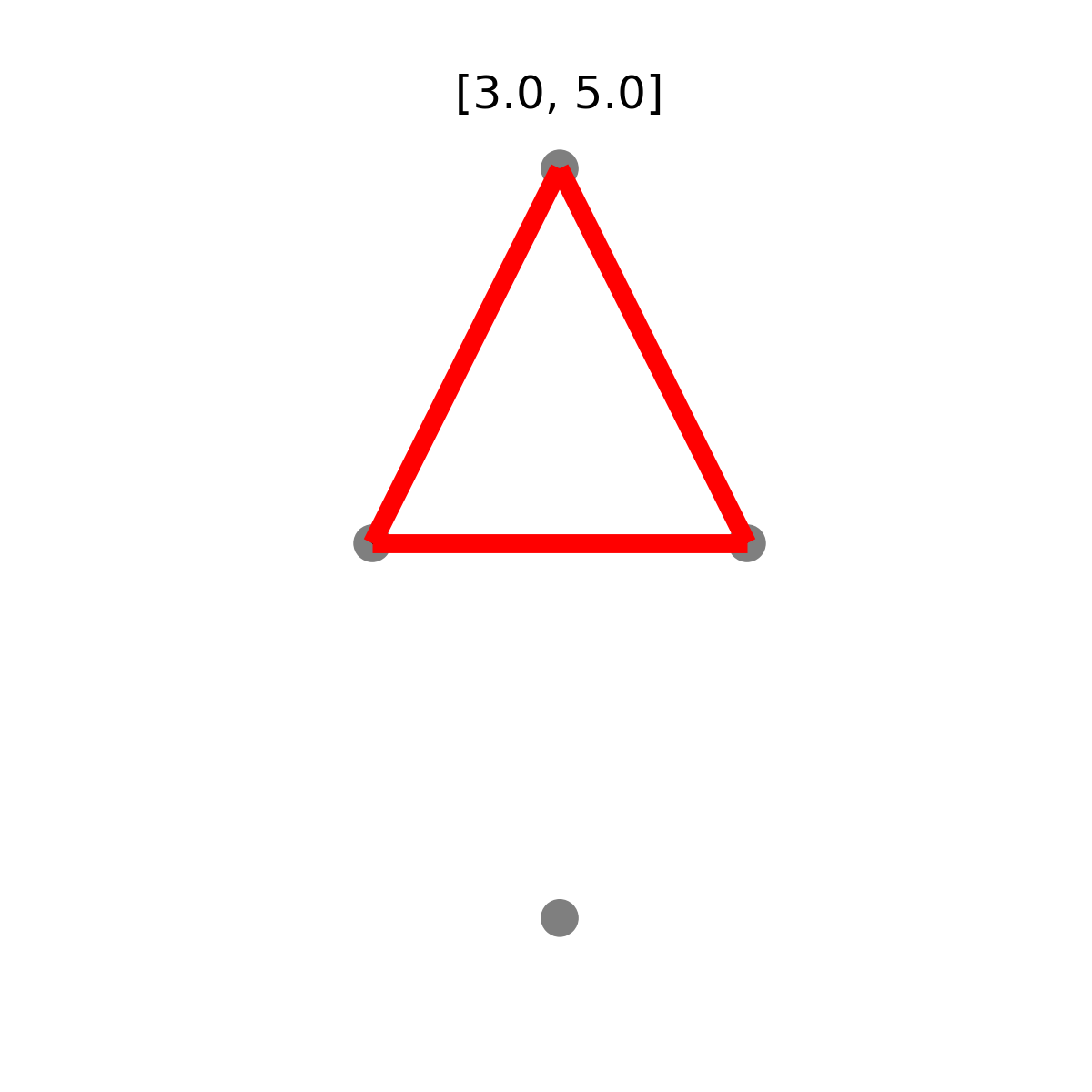}
    \subcaption{}\label{fig:harmonic_repr_example_5}
\end{subfigure}
\begin{subfigure}{.3\textwidth}
    \includegraphics[width=\textwidth]{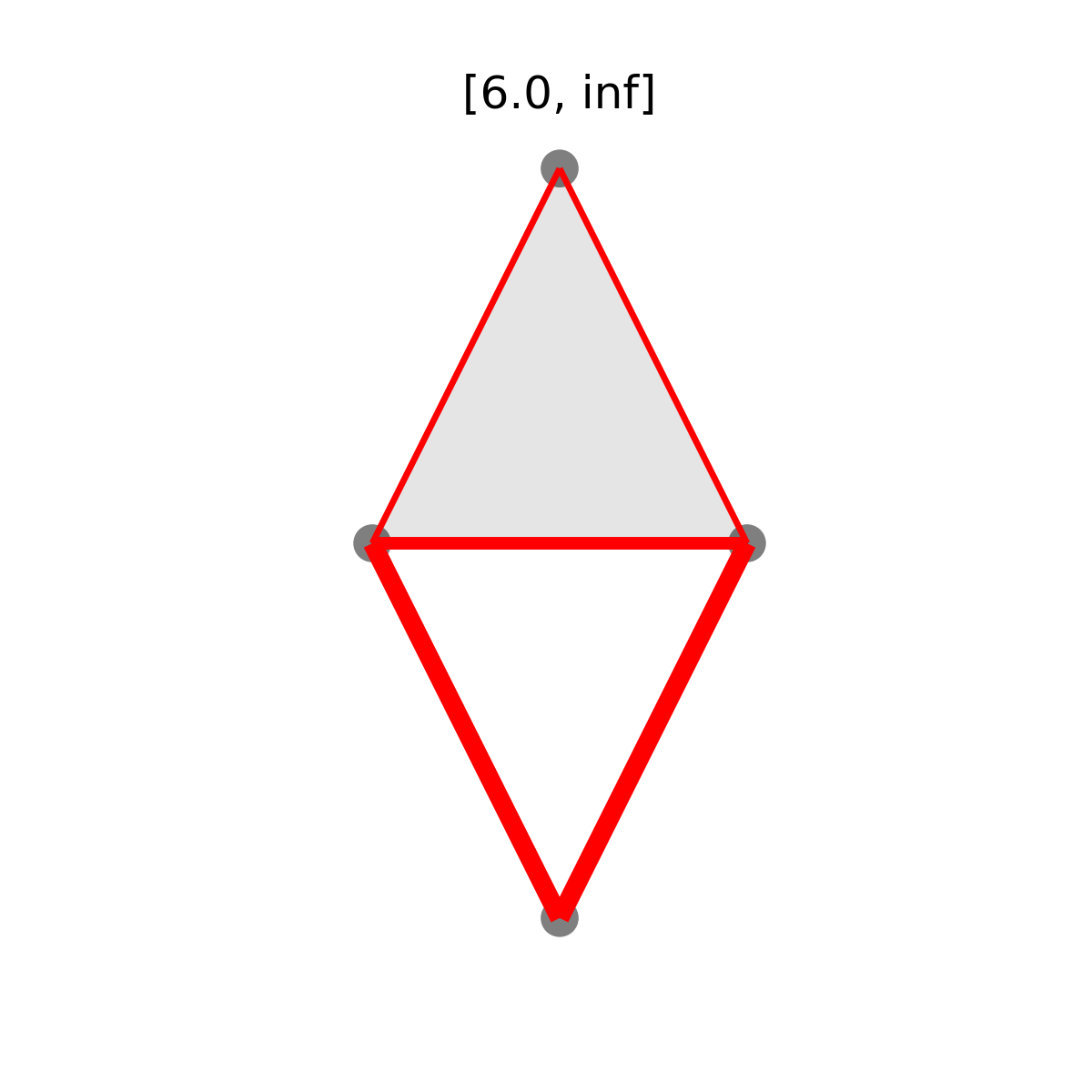}
    \subcaption{}\label{fig:harmonic_repr_example_9}
\end{subfigure}

\end{subfigure}
\caption{Example of a filtered simplicial complex and its persistent homology, found in \cite{basu2022harmonic}. Figures \ref{fig:t0} through \ref{fig:t6} illustrate the stages of the filtration, and  Figure \ref{fig:barcode_example} is the barcode associated to the filtration. Skeleton and barcode plots were created using \texttt{\textbf{maTilDA}}. Figures \ref{fig:harmonic_repr_example_5} and \ref{fig:harmonic_repr_example_9} show harmonic representatives in red, with thickness proportional to the absolute value of coefficients in a chain of a given simplex. }
\label{fig:example_filtration}
\end{figure}
%\AGS{Not sure if this matilda mention should be there, added it just in case but feel free to remove it if not ok }}\label{fig:example_filtration}

\subsection{Essential simplices and harmonic representative}
\label{subsec:essential}
Given a bar in the persistent barcode of a filtration (arising from some application) it is often natural to ask
for a cycle in the cycle space of the simplicial complex representing this bar. However, such cycles are
only defined modulo some equivalences -- and so there is no unique answer to this question. In application
(see for example \cite{basu_et_al:LIPIcs:2018:9316}) it is also important to associate a set of simplices 
to a given bar which play the most important role in the birth of the bar.
We call this set, the set of essential simplices of the bar and denote it as $\Sigma(b)$ (see \cite{basu2022harmonic} for a precise definition).

If $b$ is a bar in the barcode of a filtration and $z$ is a cycle representing $b$, then the set of simplices that  appear with a non-zero coefficient in $z$ contains 
the set of essential simplices of $b$, but could be larger. One measure of the
``quality'' of the representing cycle $z$ is the relative weights of the essential and the non-essential
simplices appearing in its support. The following quantity was defined in \cite{basu2022harmonic} as
a measure of this quality.

\begin{definition}[Relative essential content]
\label{def:content}
Let $b$ be a bar in the barcode of a filtration,
and $z = \sum_{\sigma \in K^{(p)})} c_\sigma \cdot \sigma$ is a cycle representing $z$. 
We denote
\[
\econt(z) = \left(\frac{\sum_{\sigma \in \Sigma(b)}  c_\sigma^2}{\sum_{\sigma \in K^{(p)}} c_\sigma^2}\right)^{1/2}.
\]
We will call $\econt(z)$ \emph{the relative essential content of $z$}.
\end{definition}

The following theorem proved in \cite{basu2022harmonic} and which justifies our approach in the current paper states that that the relative essential content is maximized by the harmonic representative of a bar.

\begin{theorem}[Harmonic representatives maximize relative essential content]
\label{thm:essential}
With the same notation used above, 
let $z_0$ be a harmonic representative of a bar $b$. Then for any cycle $z$ representing $b$, 
\[
\econt(z) \leq \econt(z_0).
\]
\end{theorem}

\begin{remark}
Note that Theorem~\ref{thm:essential} implies in particular that the relative essential contents
of any two harmonic representatives of a simple bar are equal. But this is clear also from the definition of the relative essential content and the fact that any two harmonic representatives of a simple bar are proportional.
\end{remark}

\subsection{Harmonic Weights from Data}
\label{subsec:harmonic_from_data}
We developed the following methodology to extract harmonic weights from multi-omics data, as depicted in Figure~\ref{fig:workflow}. Given a collection of samples together with an appropriate notion of distance (euclidean, correlation, cosine, etc.), we build the following structure on top of it and compute its persistent homology barcodes. 

\begin{definition}[Vietoris-Rips complex]
    Let $X$ be a finite collections of points.
    Given a parameter $\epsilon \geq 0$, the \emph{Vietoris-Rips complex} constructed from $X$ is the collection of all subset of diameter at most $2\epsilon$ , where the diameter is the greatest distance between any pair of vertices 
    \[
    \text{V-R}(X, \epsilon) = \{ \sigma \subseteq X \mid diam(\sigma) \leq 2\epsilon \} \quad.
    \]

    The filtration of each simplex is given by its diameter.
\end{definition} 
It should be noted that, for too large values of the parameter $\epsilon$, the size of a Vietoris-Rips complex tends to grow exponentially with the number of points. As a matter of fact, it is straightforward to show that for any $\epsilon$ larger than the largest distance between any pairs of points; the resulting complex will be fully connected and it will contain $2^n$ simplices, where $n$ is the number of data points. In order to avoid this exponential explosion, is it common practice to chose a value of $\epsilon$ strictly less than the diameter of the whole point cloud. A possible heuristic for the choice of $\epsilon$ is using the maximal distance between any point and its nearest neighbour, in order to assure the existence of only one connected component at the maximum filtration level. For the applications described in Section~\ref{sec:Results} we choose the minimal $\epsilon$ such that all $1$-dimensional bars have finite length.   

For a given dimension $p$, usually $p=1$, we select all the bars in the $p$-dimensional barcode that are longer than a threshold and compute their harmonic representatives. Such threshold can be determined as a specific quantile of the distribution of the bars' lengths. As mentioned in the previous section, each $p$-dimensional harmonic representative consist of a linear combination of $p$-simplices. In particular, each $1$-dimensional harmonic representative is a linear combination of edges. The absolute values of the coefficients of such linear combinations are the \textit{harmonic weights}. Harmonic weights take value in the unit interval $[0,1]$ and they encode how important each simplex is for a given cycle; in particular, all essential simplices have weight $1$.

Since each $p$-dimensional simplex corresponds to a collection of $(p+1)$ data points\footnote{Recall that points are $0$-dimensional simplices, edges are $1$-dimensional, etc.}, we can assign weights to the datapoints themselves by considering for each node in the Vietoris-Rips complex, the sum of the weights of its cofaces. For example, given a $1$-dimensional harmonic representative, we can translate the weights from the edges to the nodes by assigning to each node the sum of the weights of its adjacent edges. This give us a set of weights on the nodes for each bar.

\section{Results and Discussion}
\label{sec:Results}

In this section, we are going to showcase some possible applications of harmonic persistent homology in multi-omics data. We analyze a few public datasets with different omic profiles such as transcriptomics, DNA methylation, single cell RNAseq, as well as, drug resistance to highlight how our framework is able to capture intrinsic and hidden features in the data. In particular, Sections~\ref{sec:results_luad} and more analyses in Supplementary Note (Section 2.1 and 2.2) demonstrate when the weight of the edges are considered, while Sections~\ref{sec:results_tcga_multi} and~\ref{sec:results_prepost} exemplify when the weight on the nodes are considered (as described in Section~\ref{subsec:harmonic_from_data}).

\begin{figure*}[!htbp]
    \centering
    \begin{subfigure}[t]{0.5\textwidth}
        \centering
        \includegraphics[height=2.2in]{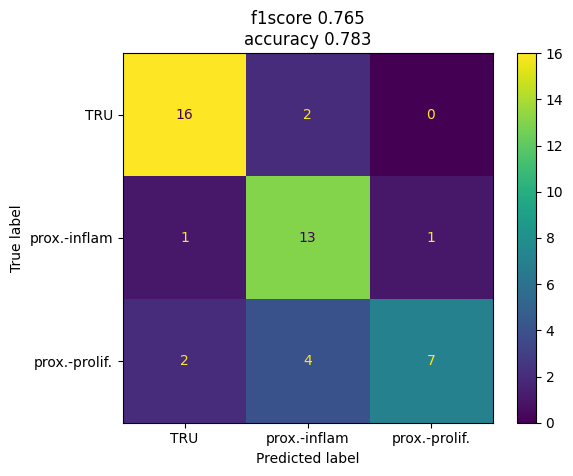}
        \caption{Standard features.}
    \end{subfigure}%
    ~ 
    \begin{subfigure}[t]{0.5\textwidth}
        \centering
        \includegraphics[height=2.2in]{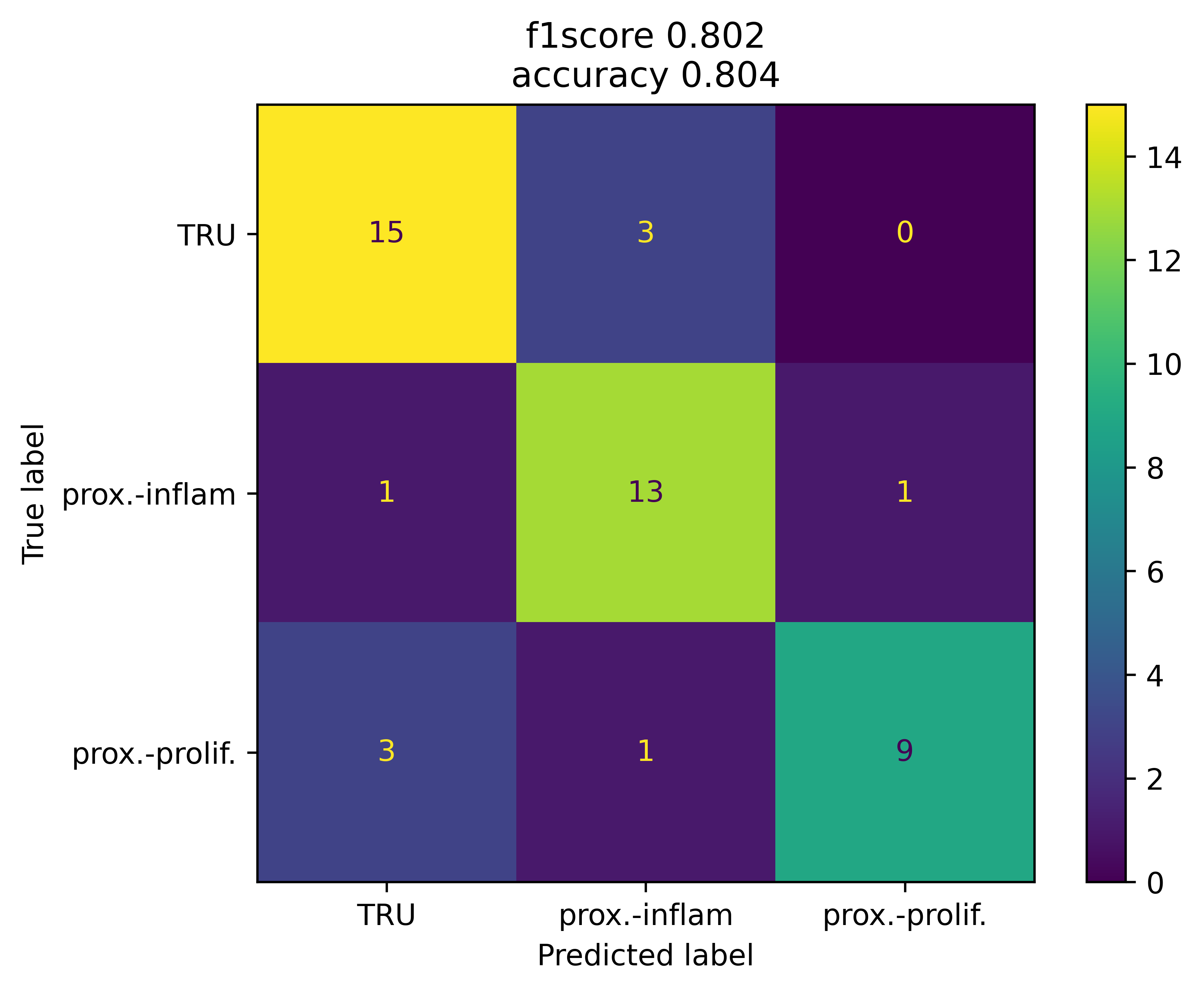}
        \caption{Harmonic weights scaled features.}
    \end{subfigure}
    \caption{Confusion matrix of the evaluation set with XGBoost classifier predicting the LUAD subtypes with the colorbar representing the number of true and false positives for each subtype.}
    \label{fig:confusion}
\end{figure*}

\subsection{Harmonic persistent homology predicts lung cancer disease subtypes}\label{sec:results_luad}
We analyzed gene expression data of $230$ samples from the Cancer Genome Atlas Lung Adenocarcinoma (TCGA-LUAD) data collection, comprising of transcriptional subtypes, such as terminal respiratory unit (TRU; $n = 89$), proximal-inflammatory (prox.-inflam; $n=78$), and proximal-proliferative (prox.-prolif; $n=63$). We selected $d = 5000$ out of 19,938 genes with highest variance in the expression levels and refer to them as \say{standard} features. We observed that the disease subtypes are clustered together when we projected the top two principal components after computing principal component analysis (PCA) on the data (Supplementary Figure 1). We then used our pipeline to enhance these features using Algorithm~\ref{alg:hph_subypes}.
\begin{algorithm}
    \caption{Disease subtype prediction with harmonic persistent homology}\label{alg:hph_subypes}
    \begin{algorithmic}[1]
    \Require A matrix, $A \in \mathbb{R}^{n \times d}$, $n$ samples, $d$ \say{standard} features.  
    \Ensure Prediction accuracy of disease subtypes 
    \State Build a Vietoris-Rips complex using $d$ features as vertex set and correlation between their expression levels across samples. 
    \State Compute the 1-dimensional persistence barcode up to a maximum filtration value of $\epsilon = 0.3$ (see Section~\ref{subsec:harmonic_from_data}). 
    \State We sorted the bars according to their birth time and selected 30 longest bars out of 100 earliest ones 
    %\DG{add reference to the early long bars paper}. 
    \State For each one of the 30 bars, compute its harmonic representative and assign a harmonic weight to each feature (see Section~\ref{subsec:harmonic_from_data}). 
    \State Obtain a single positive value for each feature by aggregating (sum) all the 30 harmonic weights for each feature. 
    \State Rescale the original features by multiplying the expression levels of each feature by its harmonic weight. 
    \State Split the data into training (70\%) and evaluation (30\%) set. 
    \State Train various machine learning models with hyperparameter tuning and five-fold cross validation on both \say{standard} features and \say{rescaled} features using harmonic weights. 
    \State Compute accuracy measures, such as the $F_1$ score, from the true positive, false positives, true negatives, and false negatives. 
    
    \end{algorithmic}
\end{algorithm}

% We built a Vietoris-Rips complex using the 5000 genes as vertex set and correlation between their expression levels across different samples as a metric. We then computed the 1-dimensional persistence barcode up to a maximum filtration value of $0.3$. Following the approach in \DG{add reference to the early long bars paper} we first sorted the bars according to their birth time and then selected the 30 longest out of the 100 earliest ones. This choice is motivated by the observation that long early bars encode significant topological information \DG{add ref}. Moreover, focusing on early bars justifies building the V-R complex only up to a maximum filtration of $0.3$ as all the considered early bars die before $0.3$. Doing so avoids the combinatorial explosion described in Section~\ref{subsec:harmonic_from_data} and allows for the computations to be executed on a consumer-grade laptop.  For each one of these 30 bars, we computed its harmonic representative and subsequently assigned an harmonic weight to each gene as described in Section~\ref{subsec:harmonic_from_data}. Finally, we summed up all the 30 harmonic weights for each gene, thus obtaining a single positive value for each gene. We then rescaled the original features by multiplying the expression levels of each gene by its harmonic weight, we will refer to this rescaled features as the \say{harmonic} features.

As part of the hyperparameter tuning process (Section 2.1 in Supplementary Note) in Algorithm~\ref{alg:hph_subypes}, 
we found XGBoost~\cite{xgboost} model to perform best on both the standard and harmonic feature sets. We observed that the $F_1$ score of the multiclass classification task on the three disease subtypes from the held-out evaluation set was 0.802 using the harmonic weights scaled features as compared to 0.765 using the \say{standard} features, which forms the baseline (Fig. ~\ref{fig:confusion}). 
We repeated the training and evaluation process 40 times with different training and held-out splits in the data and observed that the median weighted $F_1$ score on the held-out set for the classifier trained on the standard features was $0.79$, which increased to $0.81$ when training on the harmonic rescaled features. We note that only 860 of the 5,000 genes have a non-zero harmonic weight, and only 196 of them resulted in a weight greater than $0.1$. After selecting the top 20 genes sorted by their harmonic weights and performing gene set enrichment analysis we observed that adenocarcinoma of lung cancer was found as the target disease ($p < 10^{-4}$), pathways such as \textit{cellular macromolecule metabolic process} ($p < 10^{-3}$), \textit{structural constituent of ribosome} function ($p < 10^{-2}$) were found to be significant, among others. Genes such as VSIG4, DOK2, MS4A4A, CKAP5 had the highest associated harmonic weight, beyond ribosomal protein housekeeping genes such as RPS5, RPL27, etc. All of these genes are known to be associated with lung cancer~\cite{berger2013dok2,liao2014vsig4,zheng2023role} The predictive power for lung adenocarcinoma subtypes along with interpretability of genes associated with the disease with their weights as a metric of importance leads to the interpretability of standard features and a paradigm for feature selection using harmonic weights, and renders this harmonic pipeline useful for high dimensional multi-omics datasets. 
%Not only our harmonic pipeline leads to an increase in classification performance, but achieves so with a significant dimensionality reduction. \DG{not sure if we want to include this last observation.}

%\vspace*{-\baselineskip}
\subsection{Discovering breast cancer subtypes in multi-omics data}\label{sec:results_tcga_multi}
\begin{figure}[!ht]
    \centering
    \begin{subfigure}[b]{0.65\textwidth}
        \includegraphics[width=0.9\textwidth]{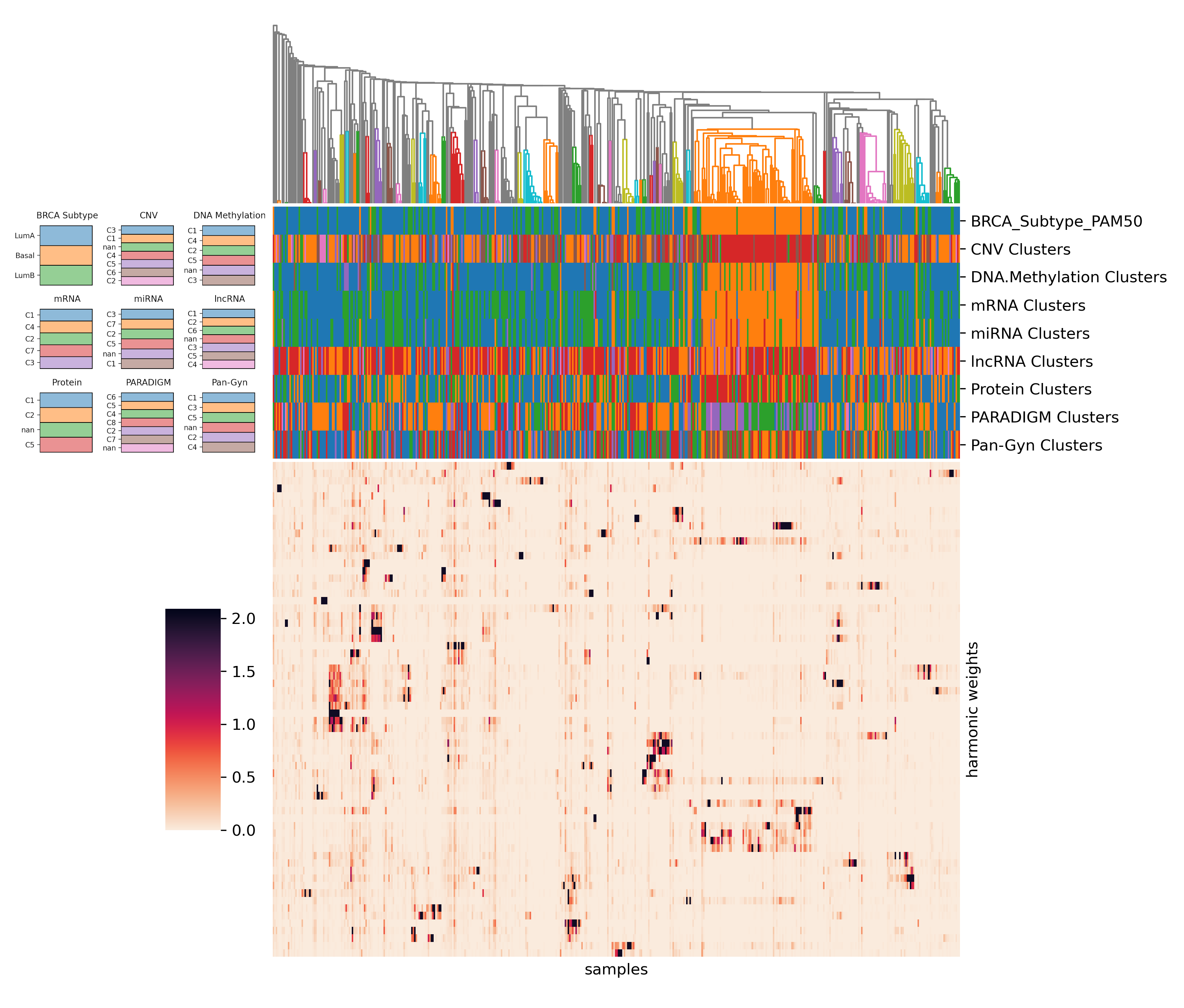}
        \caption{}
        \label{fig:clustermap_A}
    \end{subfigure}
    \hspace{-1cm}
    \begin{subfigure}[b]{0.3\textwidth}
        \includegraphics[width=\textwidth]{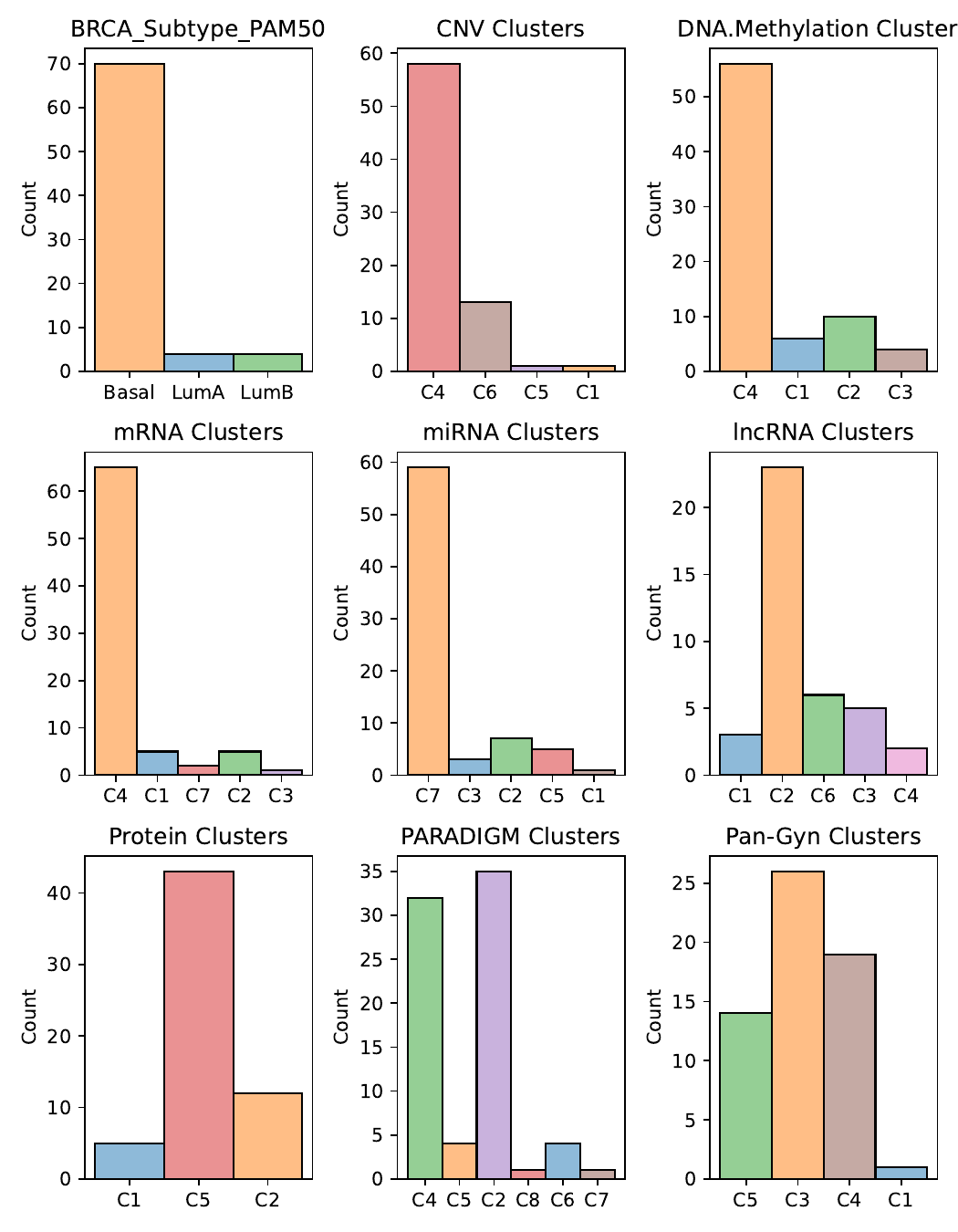}
        \caption{}
        \label{fig:clustermap_B}
    \end{subfigure}

    \caption{Single-linkage hierarchical clustering of breast cancer samples using harmonic weights is able to detect a large basal-like cluster, depicted in orange in (a). The distributions of different descriptors for samples in this cluster are shown in (b). The color scheme for each descriptor is consistent across the two panels.}
    \label{fig:clustermap}
\end{figure}

%In order to demonstrate how our framework can tackle multi-omics data we 

To showcase how harmonic persistent homology can be used in multi-omics analyses, we analyzed a set of $690$ breast cancer samples from the TCGA database for which both RNAseq and Methylation450 data are present. The dataset is comprised of $414$ Luminal-A, $141$ Luminal-B, and $135$ basal-like samples, and we considered $28,495$ genes and $363,791$ methylation sites for a total of $392,286$ features. We concatenated the RNAseq and Methylation450 data and projected them to a $100$-dimensional space using PCA. We built a Vietoris-Rips complex using distance correlation and computed its $1$-dimensional persistent homology up to a maximum filtration value of $0.75$.  We then computed the harmonic representatives of all $66$ bars longer than $0.07$ ($0.97$ quantile). For each representative we extracted the harmonic weights on the samples, resulting in a $690\times66$ weights matrix, depicted in Figure~\ref{fig:clustermap_A}. Single-linkage clustering on the samples' weights revealed a large cluster of mostly basal-like samples, whose characteristic are shown in Figure~\ref{fig:clustermap_B}. It is important to note that this cluster was found using harmonic persistent homology on samples in an unsupervised manner, without any knowledge of the nine descriptors shown on top of Figure~\ref{fig:clustermap_A}.
%relying solely on the combined RNA and methylation data. 
The clusters can be used to externally validate our findings and show how different harmonic cycles capture interactions between different subsets of data as the cycles with similar weights cluster together samples with similar descriptors. This approach can be extended to other multi-omics data, leading to a nuanced, data-driven discovery of novel subgroups of patients along with their associated biomarkers.

\vspace*{-\baselineskip}
\subsection{Transcriptomic discovery associated with treatment progression}\label{sec:results_prepost}

We consider transcriptomic data of 11 CLL patients that were treated with Venetoclax (a BCL2 inhibitor) as discussed in~\cite{10.1182/blood.2022016600} and for which RNA data were collected pre-treatment and post-treatment during which the patient disease was progressing. We selected 98 genes that are known to be associated to CLL (see Supplementary Section 3 for the full list) and provided as input to our harmonic framework the pre- and post-treatment cohorts independently.  
%to show the capability that harmonic persistent homology can  about the data with less amount information.
%It is worth pointing out that due to the limited size of the cohort classical approaches, e.g. machine learning, cannot be applied. For completeness we also consider the two cohort together and compared with a standard differential expression analysis performed in~\cite{10.1182/blood.2022016600} and found in common some important genes like NOTCH1, NFKB2 and HIST1H1B and pathways as reported later. 
\begin{figure}[!ht]
    \centering
    \includegraphics[width=0.9\linewidth]{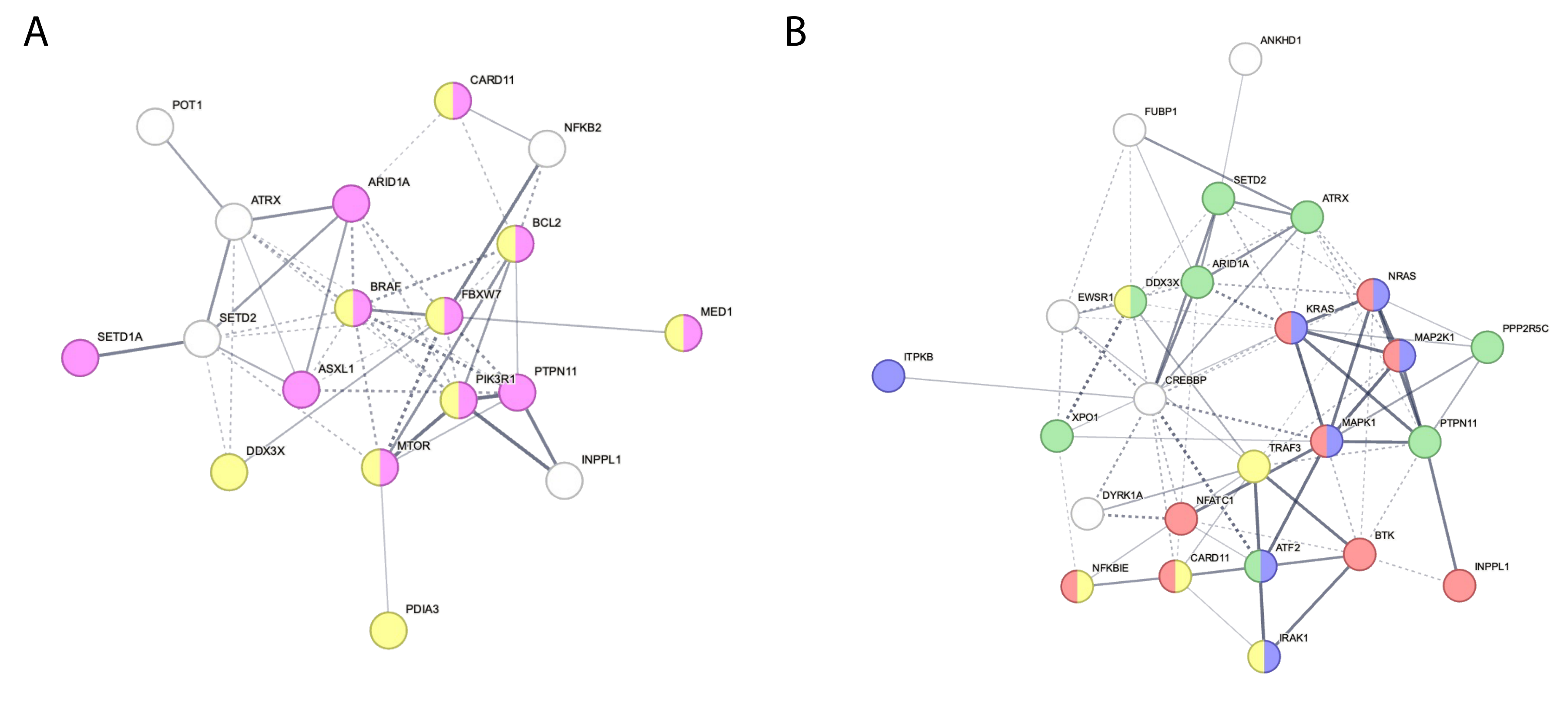}
    \caption{STRING network of the harmonic persistent homology identified genes. The edges indicate both functional and physical association. Edge thickness indicates strength of support for the edge (thick solid: >80\% of samples, thin solid: >60\%, dashed: >40\% support). A: Genes set for the pre-treatment cohort; nodes in pink belong to the Regulation of cell differentiation strength, while the one in yellow to the Reg. apoptotic process. B: Gene set for the post-treatment cohort; node belonging to MAPK cascade are in purple, 
Reg. of cell in green,
Toll like receptor cascades, and NFKappa B sig. in yellow and
BCR pathway in red.}    \label{fig:RNAprepost}
\end{figure}
Harmonic persistent homology identified relevant genes in each of the two cohorts as those genes with non-zero harmonic weights, and we performed a STRING DB network~\cite{Szklarczyk2021-lf} analysis (see~Fig~\ref{fig:RNAprepost}) to analyze protein-protein interactions of the identified genes. At pre-treatment timepoints, there was significant enrichment of genes in the apoptosis (strength 0.83, FDR 1.9e-04, pink) and regulation of cell differentiation pathways (strength 0.87, FDR 4.28e-05, yellow). In the post-treatment, progression time point, we see significant enrichment in BCR signaling (strength 1.87, FDR 2.58e-12, red) and MAPK signaling (strength 1.31, FDR 7.00e-05, purple) involving the genes \textit{BTK}, \textit{MAPK1}, \textit{KRAS}, and \textit{NRAS}. Upregulation of BCR signaling has been associated with Venetoclax progression with activation of the downstream MAPK signaling pathways~\cite{10.1182/blood.2022016600,10.1182/blood-2015-08-634816}. Furthermore, there was also enrichment of NF-$\kappa$B signaling pathways, which included the genes \textit{TRAF3} and \textit{NFKBIE} (strength 1.44, FDR 8.7e-4, yellow), and regulation of cell cycle (strength 0.85, FDR 0.0058, green). These results demonstrate the ability of our harmonic framework to extract clinically/biological plausible gene sets supported by the existing literature~\cite{10.1182/blood.2022016600,10.1182/blood-2015-08-634816} and whose clinical findings offer a measure of validation. It is worthy to note here that standard machine learning methods are not best suited to analyze this data, given the small sample size, which is often the case in various cancer datasets. However, harmonic persistent homology is capable to find previously validated and novel signals in such datasets where the sample size is very low, paving the path for efficient biomarker discovery. 
% Furthermore, highlighted and pointed out new element that could lead to new discoveries. 

\section{Conclusion}
Here we utilize the fact that harmonic cycles maximize the contribution of essential simplices, and this paper is the first application of harmonic persistent homology to biological problems. We introduce a framework that uses harmonic persistent homology to extract information from multi-omics data that enables the discovery of hidden structures in data that have the potential to inform clinical questions. Harmonic persistent homology overcomes a challenge in TDA by enabling us to systematically map the topological features uniquely to the input data and additionally, the associated weights capture the importance of the biological markers. It does so even when the data size is small, as we see in one of the applications. We applied harmonic persistent homology in a variety of scenarios in cancer with distinct questions such as subtype prediction, unsupervised subtype detection and biomarker discovery in multi-omics data.  In lung cancer, we showed that, harmonic weight rescaled features improved disease subtype prediction accuracy as compared to the baseline; in breast cancer we used harmonic weights on samples to discover a basal-like cluster; and in Venetoclax treatment response dataset, we discovered biologically relevant genes just with 11 samples. In conclusion, harmonic persistent homology has the potential of therapeutic implications for complex diseases, extending the breadth of applications of TDA in biological and healthcare data.

%as shown in our results. [@Aritra add results in term of quantitative]

% Mention that this is the first attempt to apply harmonic to biological problem. 

% We show the framework in a different scenarios, answering different fundamental questions, such as.... [something similar to the abstract as well in term of punchline - remember we are the best (somehow)]

%allows the information of the spatial structure of the data to be mapped back to the features in the data. 
\section{Author Contributions}
L.P, S.B, and A.G.S conceived this project. D.G and A.G.S implemented the harmonic persistent homology. F.U and A.B conceived the experiments and performed data quality control. All authors discussed results and wrote the manuscript.

\section{Competing Interests}
D.G., L.P., A.G.S., F.U. and A.B.  are listed (together or partially) as co-inventor of 18/506194, 18/506187, 18/616298 patent applications currently pending review at the USPTO related to the Harmonic applications.

\section{Acknowledgements}
SB was partially supported by NSF grants CCF-1910441 and CCF-2128702.
We are very grateful to Aishath Naeem for her insight in interpreting the CLL patient results, and Kahn Rhrissorrakrai for reviewing the manuscript and providing insightful comments. DG performed this work during an internship at IBM Research. 
The results shown here are in whole or part based upon data generated by the TCGA Research Network: \url{https://www.cancer.gov/tcga}.

\section{Availability and Implementation}
Source code, user manual, and sample input-output sets are available for downloading at \url{https://github.com/IBM/matilda}. 
\newpage 
\bibliographystyle{splncs04}
\bibliography{mybibliography}

\newpage

\appendix

\renewcommand{\thepage}{S\arabic{page}}
\renewcommand{\thesection}{S\arabic{section}}
\renewcommand{\thetable}{S\arabic{table}}
\renewcommand{\thefigure}{S\arabic{figure}}
\setcounter{figure}{0}
\setcounter{page}{1}

\end{document}

\bigskip
\title{Probing omics data via harmonic persistent homology}
\date{}
\author[1,$\dagger$]{Davide Gurnari}
\author[2,$\dagger$]{Aldo Guzmán-Sáenz}
\author[2]{Filippo Utro}
\author[2]{Aritra Bose}
\author[3]{Saugata Basu}
\author[2,*]{Laxmi Parida}
\affil[1]{Dioscuri Centre in Topological Data Analysis, Mathematical Institute PAN, Warsaw, PL}
\affil[2]{IBM Research, Yorktown Heights, NY }
\affil[$\dagger$]{Equal Contribution}
\affil[*]{Corresponding author: parida@us.ibm.com}

\maketitle
%\begin{linenumbers}
\input{Supplementary}
\clearpage 
\bibliographystyle{splncs04}
\bibliography{mybibliography}
%\end{linenumbers}

%% file: Supplementary.tex
\section*{Supplementary Note}
\section{An example}~\label{subsec:example}
%%We now discuss an example and calculate the harmonic homology subspaces.
%%The following example %appears in %\cite{basu2022harmonic} and is reproduced here to
We now discuss an example %and calculate the harmonic homology subspaces
to clarify the notion of harmonic representatives and essential simplices.

% \begin{figure}
%     \centering
%     \includegraphics[scale=0.65]{img/weight.pdf}
%     \caption{Example}
%     \label{fig:examplecomplex}
% \end{figure}
Let $K$ be the simplicial complex defined by (see Figure 2 at $t=6$):
\begin{eqnarray*}
K^{(0)} &=& \{[0],[1],[2],[3] \}, \\
K^{(1)} &=& \{a =[0,1], b= [1,2], c= [0,2], d = [0,3], e= [2,3] \}, \\
K^{(2)} &=& \{ t = [0,1,2] \}.
\end{eqnarray*}
%%For $p=0,1,2$, we choose the standard inner product on $C_p(K)$.
%%With these choices the matrices $M_2(K), M_1(K)$ are the following.

The matrices representing the boundary maps $\partial_2$ and $\partial_1$ are then as follows. 

\[
M_2(K) = 
\begin{blockarray}{cc}
& t \\
\begin{block}{c @{\hspace{10pt}} [c]}
a&     1  \\
b&     1 \\
c&     -1 \\
d&     0 \\
e&     0 \\
\end{block}
\end{blockarray} \quad,
\qquad
M_1(K) =
\begin{blockarray}{cccccc}
&a&b&c&d&e \\
\begin{block}{c @{\hspace{5pt}} [ccccc]}
{[0]}&     -1&  0&  -1&  -1&  0&    \\
{[1]}&     1&  -1&   0&   0&  0&    \\
{[2]}&     0&   1&   1&   0&  -1&    \\
{[3]}&     0&   0&   0&   1&  1&    \\
\end{block}
\end{blockarray}.
\]

The harmonic homology subspace (in dimension one) $\mathfrak{h}_1(K)$ is then the intersection of the null spaces
of $M_1(K)$ and $^{t} M_2(K)$, and is equal to the nullspace  of the following matrix obtained
by concatenating (row-wise) $M_1(K)$ and $^{t} M_2(K)$.

\[
\begin{blockarray}{cccccc}
&a&b&c&d&e \\
\begin{block}{c[ccccc]}
&     -1&  0&  -1&  -1&  0&    \\
&     1&  -1&   0&   0&  0&    \\
&     0&   1&   1&   0&  -1&    \\
&     0&   0&   0&   1&  1&    \\
&     1&   1&   -1&  0&  0& \\
\end{block}
\end{blockarray}.
\]

We get that $\mathfrak{h}_1(K)$ is equal to the span of the vector whose coordinates are
given by the column vector:
\[
\begin{blockarray}{cc}
& t  \\
\begin{block}{c @{\hspace{10pt}} [c]}
a  &     1  \\
b  &     1 \\
c  &     2 \\
d  &     -3 \\
e  &     3\\
\end{block}
\end{blockarray}.
\]
Notice that the absolute values of the coordinates  corresponding to the basis
vectors representing the edges $d,e$ are the greatest.
This confirms the intuition that the edges $d,e$ are more significant in the one dimensional homology 
of $K$ compared to the edges $a,b,c$. Indeed, the edges $d,e$ must occur with non-zero coefficient in 
every non-zero $1$-cycle representing a non-zero homology class in $\HH_1(K)$. The same is not true for the 
other three edges. Such edges were called \say{essential simplices} in \cite{basu_et_al:LIPIcs:2018:9316}.

\vspace*{-\baselineskip}
\section{Results}
\subsection{Lung cancer disease subtypes}
\begin{figure}
    \centering
    \includegraphics{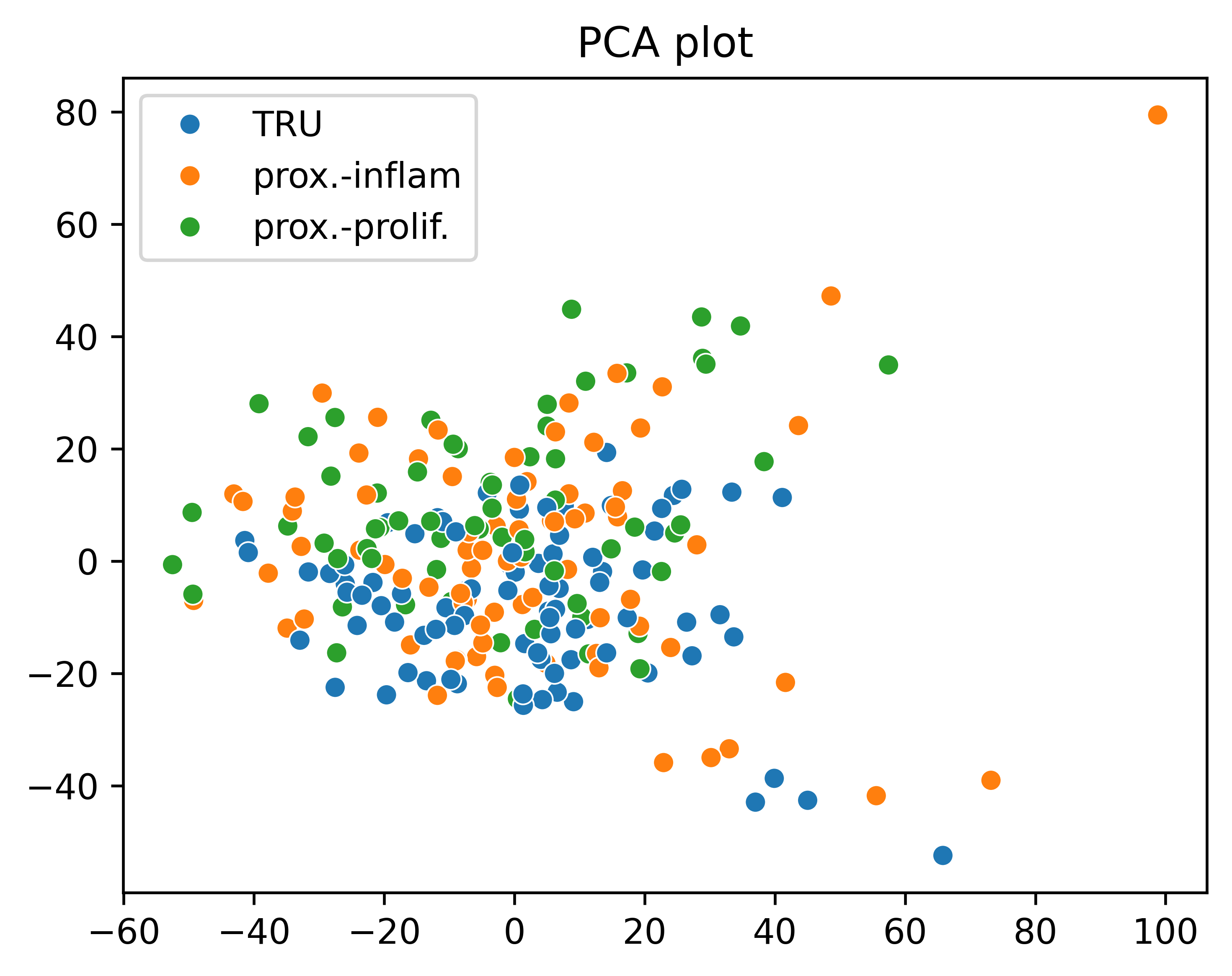}
    \caption{Top two principal components reveal the three subtypes of lung cancer data from TCGA.}
    \label{fig:pca_luad}
\end{figure}

We analyzed gene expression data of $230$ samples from the Cancer Genome Atlas Lung Adenocarcinoma (TCGA-LUAD) data collection, comprising of transcriptional subtypes such as terminal respiratory unit (TRU; $n = 89$), proximal-inflammatory (prox.-inflam; $n=78$), and proximal-profilerative (prox.-prolif; $n=63$). We selected $d = 5000$ out of 19,938 genes with highest variance in the expression levels and refer to them as \say{standard} features. We observed that the disease subtypes are clustered together when we projected the top two principal components after computing principal component analysis (PCA) on the data (Figure~\ref{fig:pca_luad}).
\subsubsection{Parameter selection}
We used a grid search with five-fold cross validation of parameters across various machine learning models such as support vector machines, logistic regression, naive bayes, random forest, AdaBoost, XGBoost, and multi-layer perceptron for the multi-class classification task of classifying the three lung cancer disease subtypes. We used scikit-learn~\cite{scikit-learn} package in Python to perform the experiments. We observed the best performing model to be XGBoost with 0.84 $F_1$ score in the training set and 0.802 in the held-out evaluation set. 

We selected the maximum filtration value $\epsilon = 0.3$ from the data as in a Vietoris-Rips complex, all 1-dimensional bars have finite length as the complex will become fully connected at some value $\epsilon_{max}$ corresponding to the maximal distance between any pair of points (Section 2.5 in the manuscript). This is true for bars of any dimensions, with the exception of the one 0-dimensional infinite bar. In theory we would like to compute the persistent homology barcode up to this maximum value to capture the full topology of the dataset. However this is often computational infeasible and a lower maximum value is selected. This might result in some bars in the output barcode having infinite length. The quantile used in the experiments was empirically selected by analyzing the distribution of the bar lengths, which in all considered examples is bimodal with a small number of bars being significantly longer then the rest. We selected the number of longest bars as 30 and 100 earliest ones after analyzing the best performing parameters in the hyperparameter tuning phase. 

\subsection{Harmonic persistent homology predicts disease subtypes - BRCA}\label{sec:results_tcga}
\begin{figure}[!h]
\centering
    \includegraphics[width=0.55\textwidth]{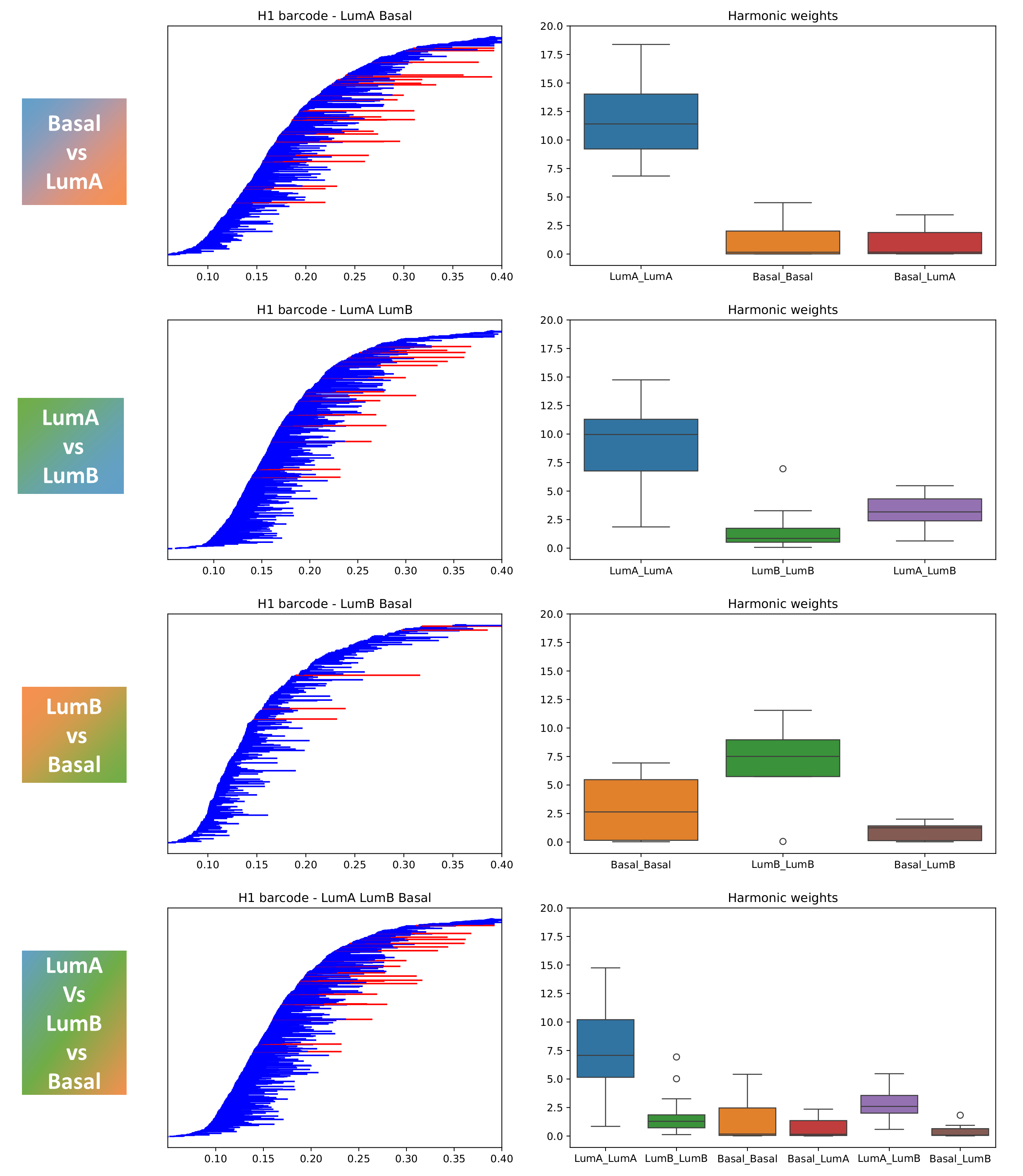}
    \caption{Harmonic persistent homology distinguishes between breast cancer subtypes (a) PCA plot, (b) 1-dimensional barcode, and (c) distribution of the cumulative harmonic weights for possible pairs of subtypes. The red colored lines in (b) indicate the bars longer than $0.08$ for which the harmonic representatives were computed. }
    \label{fig:brca_main}
\end{figure}

We analyzed gene expression data of $958$ samples from TCGA, comprising of $561$ Luminal A, $207$ Luminal B and $190$ basal-like breast cancer samples.  We then selected key genes %16 genes,
%\footnote{\textsc{GRM4, GRM8, KRT18, NMUR1, MUC1, CX3CL1, GATA3, NCAM1, CENPI, CENPK, CDC7, CCNE2, KIF18A, STIL, CDCA7, CKS2}}
%8 of them are known to be
differentially expressed for the Luminal A subtype and %the other 8 are differentially expressed for 
the basal-like subtype indicated in~\cite{jia_identification_2020}. 
For each pair of subtypes, LumA-Basal, LumB-Basal, LumA-LumB, we 
%sampled $150$ samples for each subtype \DG{now we do it for the entire dataset, no subsampling} (to ensure class balance) and 
built a Vietoris-Rips complex using distance correlation~\cite{szekely_measuring_2007} as metric between the samples, interpreted as points in a 16 dimensional gene space. We then selected all bars in the $1$-dimensional persistence barcodes having length greater than $0.08$ (threshold is determined as explained in Section~2.5) and computed their harmonic representatives. For each representative, we computed the sum of the harmonic weights for each edge, defined by the labels of its two (or three) nodes (e.g. LumA-LumA or LumA-Basal or LumA-LumB-Basal) as shown in Figure~\ref{fig:brca_main}. Our harmonic persistent homology framework is able to quantify the expected differences in gene expressions levels between samples of the same subtype as we selected key differentially expressed genes prior to computing their harmonic weights. Hence, our framework plays an instrumental role in prediction of disease subtypes.  
\vspace*{-\baselineskip}
\subsection{Discovery of disease transition using Single cell RNA data}\label{sec:results_scrna}

\begin{figure}[!ht]
    \centering
    \includegraphics[width=0.85\linewidth]{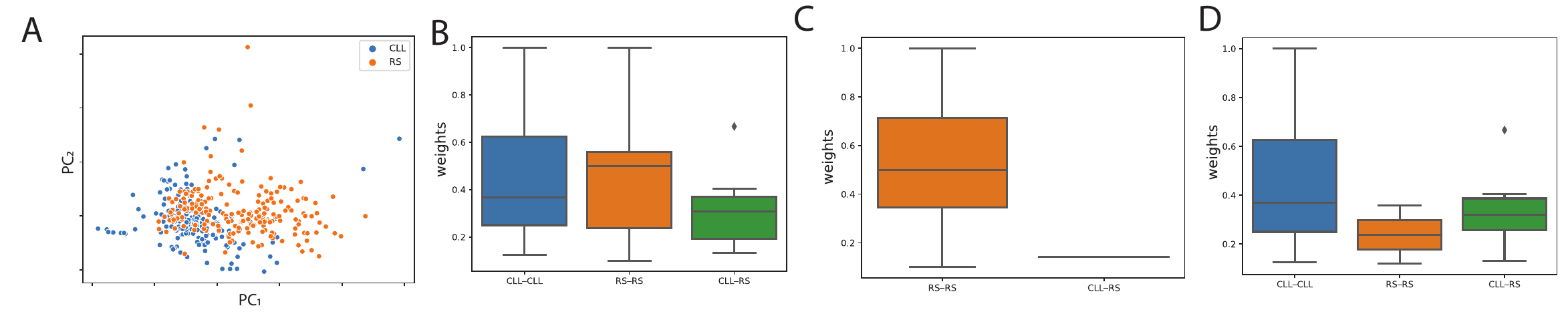}
    \caption{A. PCA of the 400 random cells from scRNA. B weights distribution of the longest cycles. C and D distribution of the weights in the first two longest cycles}
    \label{fig:scRNA}
\end{figure}

We also perform our analysis on single cell RNA data from lymphnode of a cancer under transformation to an high-grade malignancy and rapid disease progression~\cite{Parry2023-bp}. In particular,  Richter syndrome (RS), is an aggressive lymphoma developing in patients with chronic lymphocytic leukemia (CLL). The sample was analyzed with fluorescence activated cell sorting to identify CLL and RS cell. As reported in the original paper that describe the data, the sample shows a strong mixture CLL and RS cell (as shown in Fig~\ref{fig:scRNA}A). Most likely the sample was collected in an early trasformation stage. We compute the Harmonic persistent homology on a sample of 400 random cell (200 per type) in order to avoid possible class bias. Multiple runs were performed for stability of the results, and a random subset of results are shown in Fig~\ref{fig:scRNA}B-D. In particular we can see how the cycles capture the two different cell cycles. In particular, further study about of the cells involved in edges with mix label could shed some lights to the understanding of this aggressive disease and the cause of the transition from CLL to RS. 

\section{Gene list}
\label{sec:supp_genelist}
Here is the gene list used for the analysis in Section~\ref{sec:results_tcga}: \textsc{GRM4, GRM8, KRT18, NMUR1, MUC1, CX3CL1, GATA3, NCAM1, CENPI, CENPK, CDC7, CCNE2, KIF18A, STIL, CDCA7, CKS2}

\vspace{\baselineskip}

\noindent Here is the gene list used for the analysis in Section 3.3 in the manuscript:
 \textsc{GNB1,
 DNAJC11,
 MTOR,
 SPEN,
 ARID1A,
 FUBP1,
 NRAS,
 ITPKB,
 XPO1,
 ATF2,
 FSIP2,
 SF3B1,
 PLCL1,
 IRS1,
 MYD88,
 SETD2,
 KLHL6,
 FBXW7,
 TLR2,
 KIAA0947,
 ITGA1,
 ITGA2,
 SKIV2L2,
 PIK3R1,
 ANKHD1,
 IRF4,
 HIST1H1E,
 HIST1H1B,
 PIM1,
 NFKBIE,
 ZNF292,
 SYNE1,
 CARD11,
 POT1,
 C7orf55-LUC7L2,
 BRAF,
 LYN,
 CDKN2A,
 SYK,
 NOTCH1,
 TRAF2,
 EGR2,
 PLCE1,
 NFKB2,
 PLEKHA1,
 NXF1,
 INPPL1,
 BIRC3,
 ATM,
 CCND2,
 CDKN1B,
 PLCZ1,
 KRAS,
 KMT2D,
 BAZ2A,
 POLR3B,
 PTPN11,
 XPO4,
 FOS,
 PPP2R5C,
 TRAF3,
 MGA,
 PDIA3,
 MAP2K1,
 CHD2,
 CREBBP,
 SETD1A,
 NFAT5,
 PLCG2,
 ZC3H18,
 TP53,
 MED1,
 IKZF3,
 CD79B,
 BCL2,
 NFATC1,
 RPS15,
 CD79A,
 BAX,
 CNOT3,
 ASXL1,
 SAMHD1,
 DYRK1A,
 MAPK1,
 IGLL5,
 CHEK2,
 EWSR1,
 BCOR,
 DDX3X,
 MED12,
 ZMYM3,
 ATRX,
 BTK,
 NKAP,
 ELF4,
 IRAK1,
 FAM50A,
 BRCC3}